\documentclass[journal]{IEEEtran}
\IEEEoverridecommandlockouts
\usepackage{cite}
\usepackage{amsmath,amssymb,amsfonts}
\usepackage{graphicx}
\usepackage{acronym,comment}
\usepackage[colorinlistoftodos]{todonotes}

\usepackage{textcomp}
\usepackage{xcolor}

\usepackage{bbm}
\usepackage{amsthm}
\usepackage{psfrag,caption,subcaption}

\usepackage{bbm}
\usepackage{algorithm}
\usepackage[noend]{algpseudocode} % algorithm

\algrenewcommand\algorithmiccomment[1]{// {\itshape #1}}
\usepackage[utf8]{inputenc}
\usepackage[english]{babel}
\newtheorem{theorem}{Theorem}

\def\cast{{
   \mathord{
      \hbox to 0em{
         \ooalign{
	   \smash{\hbox{$\ast$}}\crcr
	   \smash{\hskip-1pt\Large\hbox{$\circ$}} }
	 \hidewidth}
      \phantom{\bigcirc}
} }}
\newcommand{\rH}{^{ \raisebox{1pt}{$\rm \scriptscriptstyle H$}}}
\newcommand{\rT}{^{ \raisebox{1.2pt}{$\rm \scriptstyle T$}}}

\newcommand{\bds}{\begin {itemize}}
\newcommand{\eds}{\end {itemize}}
\newcommand{\bdf}{\begin{definition}}
\newcommand{\blm}{\begin{lemma}}
\newcommand{\edf}{\end{definition}}
\newcommand{\elm}{\end{lemma}}
\newcommand{\bthm}{\begin{theorem}}
\newcommand{\ethm}{\end{theorem}}
\newcommand{\bprp}{\begin{prop}}
\newcommand{\eprp}{\end{prop}}
\newcommand{\bcl}{\begin{claim}}
\newcommand{\ecl}{\end{claim}}
\newcommand{\bcr}{\begin{coro}}
\newcommand{\ecr}{\end{coro}}
\newcommand{\bquest}{\begin{question}}
\newcommand{\equest}{\end{question}}

\newcommand{\larrow}{{\larrow}}

\newcommand{\argmin}{\ensuremath{\mathrm{arg}\min}}
\newcommand{\argmax}{\ensuremath{\mathrm{arg}\max}}

\newcommand{\cO}{{\ensuremath{\mathcal{O}}}}
\newcommand{\cP}{{\ensuremath{\mathcal{P}}}}

\def\mbC{{\ensuremath{\mathbb C}}}

\def\mbE{{\ensuremath{\mathbb E}}}

\newcommand{\va}{{\ensuremath{{\mathbf{a}}}}}

\newcommand{\vb}{{\ensuremath{{\mathbf{b}}}}}
\newcommand{\vc}{{\ensuremath{{\mathbf{c}}}}}
\newcommand{\vd}{{\ensuremath{{\mathbf{d}}}}}

\newcommand{\vg}{{\ensuremath{{\mathbf{g}}}}}

\newcommand{\vh}{{\ensuremath{{\mathbf{h}}}}}

\newcommand{\vr}{{\ensuremath{{\mathbf{r}}}}}
\newcommand{\vs}{{\ensuremath{{\mathbf{s}}}}}

\newcommand{\vu}{{\ensuremath{{\mathbf{u}}}}}

\newcommand{\vw}{{\ensuremath{{\mathbf{w}}}}}

\newcommand{\vx}{{\ensuremath{{\mathbf{x}}}}}

\newcommand{\vz}{{\ensuremath{{\mathbf{z}}}}}

\newcommand{\mA}{{\ensuremath{\mathbf{A}}}}

\newcommand{\mB}{{\ensuremath{\mathbf{B}}}}

\newcommand{\mC}{{\ensuremath{\mathbf{C}}}}

\newcommand{\mD}{{\ensuremath{\mathbf{D}}}}

\newcommand{\mG}{{\ensuremath{\mathbf{G}}}}
\newcommand{\mH}{{\ensuremath{\mathbf{H}}}}

\newcommand{\mI}{{\ensuremath{\mathbf{I}}}}

\newcommand{\mQ}{{\ensuremath{\mathbf{Q}}}}

\newcommand{\mR}{{\ensuremath{\mathbf{R}}}}

\newcommand{\mS}{{\ensuremath{\mathbf{S}}}}

\newcommand{\mU}{{\ensuremath{\mathbf{U}}}}

\newcommand{\mW}{{\ensuremath{\mathbf{W}}}}

\usepackage{latexsym}

\def\IC{\mathbb C}
\def\IN{\mathbb N}
\def\IZ{\mathbb Z}
\def\IR{\mathbb R}

\def\shat{^{\mathchoice{}{}%
 {\,\,\smash{\hbox{\lower4pt\hbox{$\widehat{\null}$}}}}%
 {\,\smash{\hbox{\lower3pt\hbox{$\hat{\null}$}}}}}}

\def\bSigma{{
      \ooalign{
      \smash{\hskip.4pt\raise.4pt\hbox{$\Sigma$}}\vphantom{}\crcr
      \smash{\hskip.7pt\raise.6pt\hbox{$\Sigma$}}\vphantom{}\crcr
      \smash{\hbox{$\Sigma$}}\vphantom{$\Sigma$}}
      \vphantom{\hbox{$\Sigma$}}
      }}
\def\bTheta{{
      \ooalign{
      \smash{\hskip.5pt\raise.5pt\hbox{$\Theta$}}\vphantom{}\crcr
      \smash{\hskip.0pt\raise.1pt\hbox{$\Theta$}}\vphantom{}\crcr
      \smash{\hbox{$\Theta$}}\vphantom{$\Theta$}}
      \vphantom{\hbox{$\Theta$}}
      }}
\def\bDelta{{
      \ooalign{
      \smash{\hskip.4pt\raise.4pt\hbox{$\Delta$}}\vphantom{}\crcr
      \smash{\hskip.7pt\raise.6pt\hbox{$\Delta$}}\vphantom{}\crcr
      \smash{\hbox{$\Delta$}}\vphantom{$\Delta$}}
      \vphantom{\hbox{$\Delta$}}
      }}
\def\bLambda{{
      \ooalign{
      \smash{\hskip.5pt\raise.5pt\hbox{$\Lambda$}}\vphantom{}\crcr
      \smash{\hskip.0pt\raise.1pt\hbox{$\Lambda$}}\vphantom{}\crcr
      \smash{\hbox{$\Lambda$}}\vphantom{$\Lambda$}}
      \vphantom{\hbox{$\Lambda$}}
      }}

\makeatletter

\def\bordermatrix#1{\begingroup \m@th
  \@tempdima 8.75\p@
  \setbox\z@\vbox{%
    \def\cr{\crcr\noalign{\kern2\p@\global\let\cr\endline}}%
    \ialign{$##$\hfil\kern2\p@\kern\@tempdima&\thinspace\hfil$##$\hfil
      &&\quad\hfil$##$\hfil\crcr
      \omit\strut\hfil\crcr\noalign{\kern-\baselineskip}%
      #1\crcr\omit\strut\cr}}%
  \setbox\tw@\vbox{\unvcopy\z@\global\setbox\@ne\lastbox}%
  \setbox\tw@\hbox{\unhbox\@ne\unskip\global\setbox\@ne\lastbox}%
  \setbox\tw@\hbox{$\kern\wd\@ne\kern-\@tempdima\left[\kern-\wd\@ne
    \global\setbox\@ne\vbox{\box\@ne\kern2\p@}%
    \vcenter{\kern-\ht\@ne\unvbox\z@\kern-\baselineskip}\,\right]$}%
  \null\;\vbox{\kern\ht\@ne\box\tw@}\endgroup}
\makeatother

\makeatletter
\def\argmin{\mathop{\operator@font arg\,min}}
\def\argmax{\mathop{\operator@font arg\,max}}
\makeatother

\newcommand{\bea}{\begin{array}}
\newcommand{\ena}{\end{array}}
\newcommand{\beq}{\begin{equation}}
\newcommand{\enq}{\end{equation}}

\newcommand{\beqa}{\begin{eqnarray}}
\newcommand{\enqa}{\end{eqnarray}}

\newcommand{\beqan}{\begin{eqnarray*}}
\newcommand{\enqan}{\end{eqnarray*}}

\newcommand{\AL}{\begin{enumerate}}
\newcommand{\ALE}{\end{enumerate}}

\def\addots{\mathinner{
    \mkern1mu\raise0pt\vbox{\kern7pt\hbox{.}}
    \mkern2mu\raise4pt\hbox{.}
    \mkern2mu\raise7pt\hbox{.}
    \mkern1mu}}

\def\sddots{\mathinner{
    \mkern.8mu\raise7pt\hbox{.}
    \mkern.8mu\raise4pt\hbox{.}
    \mkern.8mu\raise0pt\vbox{\kern7pt\hbox{.}}
    \mkern1mu}}

\def\saddots{\mathinner{
    \mkern.2mu\raise0pt\vbox{\kern7pt\hbox{.}}
    \mkern.2mu\raise4pt\hbox{.}
    \mkern.2mu\raise7pt\hbox{.}
    \mkern1mu}}

\def\sqplus{\mathbin{
	{\ooalign{\hfil\raise.3ex\hbox{\scriptsize
	+}\hfil\crcr\mathhexbox274\crcr\mathhexbox275}}
	}} 
\def\sqminus{\mathbin{
	{\ooalign{\hfil\raise.3ex\hbox{\scriptsize
	--}\hfil\crcr\mathhexbox274\crcr\mathhexbox275}}
	}}

\def\IC{{
   \mathord{
      \hbox to 0em{
	 \hskip-4pt
         \ooalign{
	   \smash{\hskip1.9pt\raise2.6pt\hbox{$\scriptscriptstyle |$}}\crcr
	   \smash{\hbox{\rm\sf C}} }
	 \hidewidth}
      \phantom{\hbox{\rm\sf C}}
} }}
\def\IN{
    {\ooalign{
   \smash{\hskip2.2pt\raise1.5pt\hbox{$\scriptscriptstyle |$}}\vphantom{}\crcr
   \hbox{\sf N}
	}}
	} 
\def\IZ{
    {\ooalign{
   \smash{\hskip1.9pt\raise0pt\hbox{$\sf Z$}}\vphantom{}\crcr
   \hbox{\sf Z}
	}}
	} 
\def\IR{
    {\ooalign{
   \smash{\hskip2.2pt\raise1.5pt\hbox{$\scriptscriptstyle |$}}\vphantom{}\crcr
   \smash{\hskip2.2pt\raise3.3pt\hbox{$\scriptscriptstyle |$}}\vphantom{}\crcr
   \hbox{\sf R}
	}}
	} 

\DeclareMathAlphabet{\mathcmb}{OT1}{cmr}{b}{n}

\def\bSigma{\ensuremath{\mathcmb{\Sigma}}}
\def\bLambda{\ensuremath{\mathcmb{\Lambda}}}

\def\bTheta{\ensuremath{\mathcmb{\Theta}}}

\newcommand{\SI}{\begin{indlist}}
\newcommand{\EI}{\end{indlist}}

\newcommand{\DL}{\begin{dashlist}}
\newcommand{\DLE}{\end{dashlist}}

\makeatletter
\def\setboxz@h{\setbox\z@\hbox}
\def\wdz@{\wd\z@}
\def\boxz@{\box\z@}
\def\underset#1#2{\binrel@{#2}%
  \binrel@@{\mathop{\kern\z@#2}\limits_{#1}}}
\def\binrel@#1{\begingroup
  \setboxz@h{\thinmuskip0mu
    \medmuskip\m@ne mu\thickmuskip\@ne mu
    \setbox\tw@\hbox{$#1\m@th$}\kern-\wd\tw@
    ${}#1{}\m@th$}%
  \edef\@tempa{\endgroup\let\noexpand\binrel@@
    \ifdim\wdz@<\z@ \mathbin
    \else\ifdim\wdz@>\z@ \mathrel
    \else \relax\fi\fi}%
  \@tempa
}
\let\binrel@@\relax%
\makeatother

\acrodef{ris}[RIS]{reconfigurable intelligent surface}
\acrodef{isac}[ISAC]{integrated sensing and communication}
\acrodef{dfbs}[DFBS]{Dual function radar communication base station}
\acrodef{ue}[UE]{user equipment}
\acrodef{sinr}[SINR]{signal-to-interference-plus-noise ratio}
\acrodef{snr}[SNR]{signal-to-noise ratio}
\acrodef{mui}[MUI]{multi-user-interference}
\acrodef{mimo}[MIMO]{multiple input multiple output}
\acrodef{ura}[URA]{uniform rectangular array}
\acrodef{nlos}[NLoS]{non-line-of-sight}
\acrodef{los}[LoS]{line-of-sight}
\acrodef{wrt}[w.r.t.]{with respect to}
\acrodef{rcc}[RCC]{radar-communication-coexistence}
\acrodef{ula}[ULA]{uniform linear array}
\acrodef{sdp}[SDP]{semi definite relaxation}
\def\BibTeX{{\rm B\kern-.05em{\sc i\kern-.025em b}\kern-.08em
		T\kern-.1667em\lower.7ex\hbox{E}\kern-.125emX}}
\begin{document}
	
	\title{Beamforming in Integrated Sensing and Communication Systems with Reconfigurable Intelligent Surfaces} 
		\author{R.S. Prasobh Sankar,~\IEEEmembership{Student Member, IEEE,} Sundeep Prabhakar Chepuri,~\IEEEmembership{Member, IEEE}, and Yonina~C.~Eldar,~\IEEEmembership{Fellow,~IEEE}
		\thanks{

			R.S.P. Sankar and S.P. Chepuri  are with the Department of Electrical Communication Engineering, Indian Institute of Science, Bangalore, India. Email:\{prasobhr,spchepuri\}@iisc.ac.in. 
			
			Y.C.~Eldar is with the Department of Mathematics and Computer Science, Weizmann Institute of Science, Rehovot, Israel. Email:yonina.eldar@weizmann.ac.il. 
			
				This work is supported in part by the Next Generation Wireless Research and Standardization on 5G and Beyond project, MeitY, Govt. of India, and Prime Minister’s Research Fellowship (PMRF), Govt. of India.

	}}

	\maketitle

	\begin{abstract}
		We consider transmit beamforming and reflection pattern design in reconfigurable intelligent surface (RIS)-assisted \ac{isac} systems to jointly precode communication symbols and radar waveforms. We treat two settings of multiple users and targets. In the first, we use a single RIS to enhance the communication performance of the ISAC system and design beams with  good cross-correlation properties to match a desired beam pattern while guaranteeing a desired signal-to-interference plus noise ratio (SINR) for each user. In the second setting, we use two dedicated RISs  to aid the ISAC system, wherein the beams are designed to maximize the worst-case target illumination power while guaranteeing a desired SINR for each user. We propose solvers based on alternating optimization as the design problems in both cases are non-convex optimization problems. Through a number of numerical simulations, we demonstrate the advantages of RIS-assisted ISAC systems. In particular, we show that the proposed single-RIS assisted ISAC system improves the minimum user SINR while suffering from a moderate loss in radar target illumination power. On the other hand, the dual-RIS assisted ISAC system improves both minimum user SINR as well as worst-case target illumination power at the targets, especially when the users and targets are not directly visible.

	\end{abstract}
	
	
	\section{Introduction} \label{sec:intro}

		\IEEEPARstart{I}{ntegrated} sensing and communication (ISAC) systems are envisioned to play a crucial role in 5G advanced and 6G wireless networks~\cite{rajatheva2020white,liu2022ISAC_6G,wymeersch2021Integration_comm_sense_6G,nemati2022towardsJRC}. \ac{dfbs} is an example of an \ac{isac} system that aims to establish communications between \acp{ue} while using the same resources to carry out sensing tasks~\cite{mishra2019towards_mmWave,ma2020jrc_autonomous_vehicle,chiriyath2017radar_commun}. The integration of communication and sensing functionalities is usually achieved by using communication waveforms to also carry out radar sensing, embedding communication symbols in radar waveforms, or designing  precoders to jointly transmit communication and sensing waveforms~\cite{zheng2019radar_comm,liu2020joint_radar}.
	
		One of the main drawbacks of using radar (communication) signals to carry out both communication and sensing is the inevitable degradation in the communication (respectively, radar) performance. For example, using radar waveforms for \acs{isac} limits the communication data rate to the order of the pulse repetition interval of the radar waveform. This issue can be alleviated by using dedicated beamformers to carry out communication and sensing functionalities~\cite{liu2020joint_transmit_beamform,liu2021cramer_rao_isac,pritzker2022transmit_precoder}. Furthermore, even though dedicated beams are used for sensing, \ac{dfbs} can exploit communication waveforms for radar sensing as well without treating it as interference as it has full knowledge of both waveforms.

		 Transmit beamformers in \ac{mimo} radar systems are typically designed to either  achieve a desired beampattern at the transmitter or to maximize the illumination power at different target directions~\cite{stoica2007on_probing_signal}. 
			The design of transmit beamformers in~\cite{stoica2007on_probing_signal} has been extended to ISAC systems in~\cite{liu2020joint_transmit_beamform,liu2021cramer_rao_isac}, where different beamformers for \acp{ue} and target directions are designed to guarantee a minimum \ac{sinr} to the \acp{ue} while matching a desired beampattern at the transmitter~\cite{liu2020joint_transmit_beamform} or to minimize the radar sensing Cram\'er-Rao lower bound~\cite{liu2021cramer_rao_isac}.
		    In~\cite{pritzker2022transmit_precoder}, the authors propose a radar priority approach, wherein the radar~\ac{snr} is maximized while serving as many users as possible, each with a minimum~\ac{sinr}. 
			
			Large available bandwidths at mmWave frequency bands can be exploited to achieve high data rates and to obtain improved range resolution for radar sensing. Operating at mmWave frequencies is challenging due to the pathloss, which is so severe that the \ac{nlos} paths can be too weak to be of any practical use, preventing reliable communication or sensing.  However, when the direct links to the \acp{ue} or targets are weak or absent, the aforementioned methods~\cite{liu2020joint_transmit_beamform,pritzker2022transmit_precoder,liu2021cramer_rao_isac} will not be able to provide desired levels of radar or communication performance. To combat such harsh propagation environments, an emerging technology with large two-dimensional array of passive reflectors, referred to as \acp{ris} is receiving significant attention, separately for wireless communications~\cite{basar2019wireless,renzo2020smart}, localization~\cite{wymeersch2020beyond_5g_localization}, and radar sensing~\cite{buzzi2022foundations,wang2021ris_radar,lu2021ris_mimo_radar,shao2022target_sensing}. 
	
		\ac{ris} is a two-dimensional array of phase shifters, which can be independently tuned from a controller often at the base station. Although \acp{ris} do not have any signal processing capabilities to perform data acquisition, channel estimation, or symbol decoding, its phase profile can be designed to favorably alter the wireless propagation environment and introduce \textit{virtual} \ac{los} paths, which enable communication or sensing even when the direct path is weak or blocked.

		Integrating \ac{ris} into ISAC has been considered recently in several works~\cite{jiang2021irs_dfrc,he2022risassisted,song2022JRC_RIS,sankar2021mmWave_JRC,liu2022joint_passive,li2022dual,liu2022joint,wang2021joint_waveform_design,wang2022joint_waveform_discrete,zhu2022intelligent,shtaiwi2021sum_rate_RIS_ISAC,hua2022joint} to achieve improved per user \ac{sinr} and/or sum target illumination power, especially when the direct path from the \ac{dfbs} to the \acp{ue} or targets are blocked. 
		 Typically, the design of transmit beamformers~(or  waveforms) and the \ac{ris} phase shifts is carried out  to achieve desired levels of communication and sensing performance. Usual choices of communication performance metric are \ac{snr}~\cite{jiang2021irs_dfrc,he2022risassisted,song2022JRC_RIS,li2022dual}, \ac{sinr}~\cite{zhu2022intelligent,hua2022joint}, sum-rate~\cite{liu2022joint_passive,shtaiwi2021sum_rate_RIS_ISAC}, or multi-user interference~\cite{wang2021joint_waveform_design,wang2022joint_waveform_discrete,liu2022joint}. Commonly used radar performance metrics are target received \ac{snr}~\cite{jiang2021irs_dfrc,he2022risassisted,liu2022joint_passive,li2022dual,liu2022joint},  sum-\ac{snr}~\cite{hua2022joint}, worst-case target illumination power~\cite{song2022JRC_RIS}, transmit beampattern mismatch error~\cite{wang2021joint_waveform_design,zhu2022intelligent}, direction of arrival estimation Cram\'er-Rao bound~\cite{wang2022joint_waveform_discrete}, or the interference of communication waveforms on radar~\cite{shtaiwi2021sum_rate_RIS_ISAC}. Most of the aforementioned  works  consider  relatively simple settings such as single user multiple targets~\cite{song2022JRC_RIS},  multiple users single target~\cite{liu2022joint,liu2022joint_passive,li2022dual}, or  using only single-antenna for communications~\cite{he2022risassisted}. Methods designed to work with single \ac{ue} often use \ac{snr} as the communication metric. However, in the presence of multiple users, it is typical to use  SINR as a metric, which is a fractional function that cannot be readily handled using the solvers in~\cite{jiang2021irs_dfrc,he2022risassisted,song2022JRC_RIS}.  Similarly, methods intended to sense only a single target such as~\cite{jiang2021irs_dfrc,liu2022joint_passive,li2022dual,liu2022joint} use radar \ac{snr} as the metric. In presence of multiple targets, metrics such as worst-case target illumination power, beampattern mismatch error, or sum-radar \ac{snr} is preferred. Hence algorithms designed for handling a single target~\cite{jiang2021irs_dfrc,liu2022joint_passive,li2022dual,liu2022joint} cannot be readily extended to sense multiple targets.

		The most general setting of RIS-assisted \ac{isac} systems for multiple users and targets is considered in~\cite{zhu2022intelligent,shtaiwi2021sum_rate_RIS_ISAC,wang2021joint_waveform_design,wang2022joint_waveform_discrete,hua2022joint}. In~\cite{wang2021joint_waveform_design,wang2022joint_waveform_discrete}, a single dedicated transmit waveform is used  for communication and sensing. 	In~\cite{zhu2022intelligent}, transmit beamformers and reflection coefficients are designed for an \ac{isac} system having separate colocated subarrays for sensing and communication. Due to the close proximity of radar and communication subarrays, the signal received at the \ac{ris} will also receive a significant amount of radar waveforms, resulting in increased interference at the \acp{ue}, that is however ignored in~\cite{zhu2022intelligent}.  Moreover, target sensing is not possible whenever the direct paths to the targets are blocked since the targets are directly sensed by the \ac{dfbs} without any aid from the \ac{ris}. 
			
			An \ac{ris} assisted \ac{rcc} system is considered in~\cite{shtaiwi2021sum_rate_RIS_ISAC}, where geographically separated transceivers are used for sensing and communication. However, the design of radar waveforms to achieve a certain radar performance is not considered in~\cite{shtaiwi2021sum_rate_RIS_ISAC} with the study being limited to the reduction of communication interference on the radar system. 
		 In~\cite{hua2022joint}, the sum radar \ac{snr} due to multiple targets is used as the radar performance metric to design separate communication and sensing beamformers. Considering the sum radar \ac{snr} as a radar metric leads to scenarios with the entire power being transmitted towards one of the targets, resulting in the radar system completely missing one or more targets~(see Fig.~\ref{fig:alg1_beams}(a) and \ref{fig:alg1_beams}(b) in Section~\ref{sec:simulations:comm_ris}). 
		Furthermore, \cite{zhu2022intelligent,shtaiwi2021sum_rate_RIS_ISAC,hua2022joint,wang2021joint_waveform_design,wang2022joint_waveform_discrete} consider a setting where the \ac{ris} is not used for radar sensing. Hence, whenever the targets are not directly visible to the \ac{dfbs}, it is not possible to leverage the benefit of \ac{ris} to introduce additional paths that enable reliable sensing~\cite{buzzi2022foundations}. 
	
		\begin{figure*}[ht]
		\centering
		\includegraphics[width=0.9\linewidth]{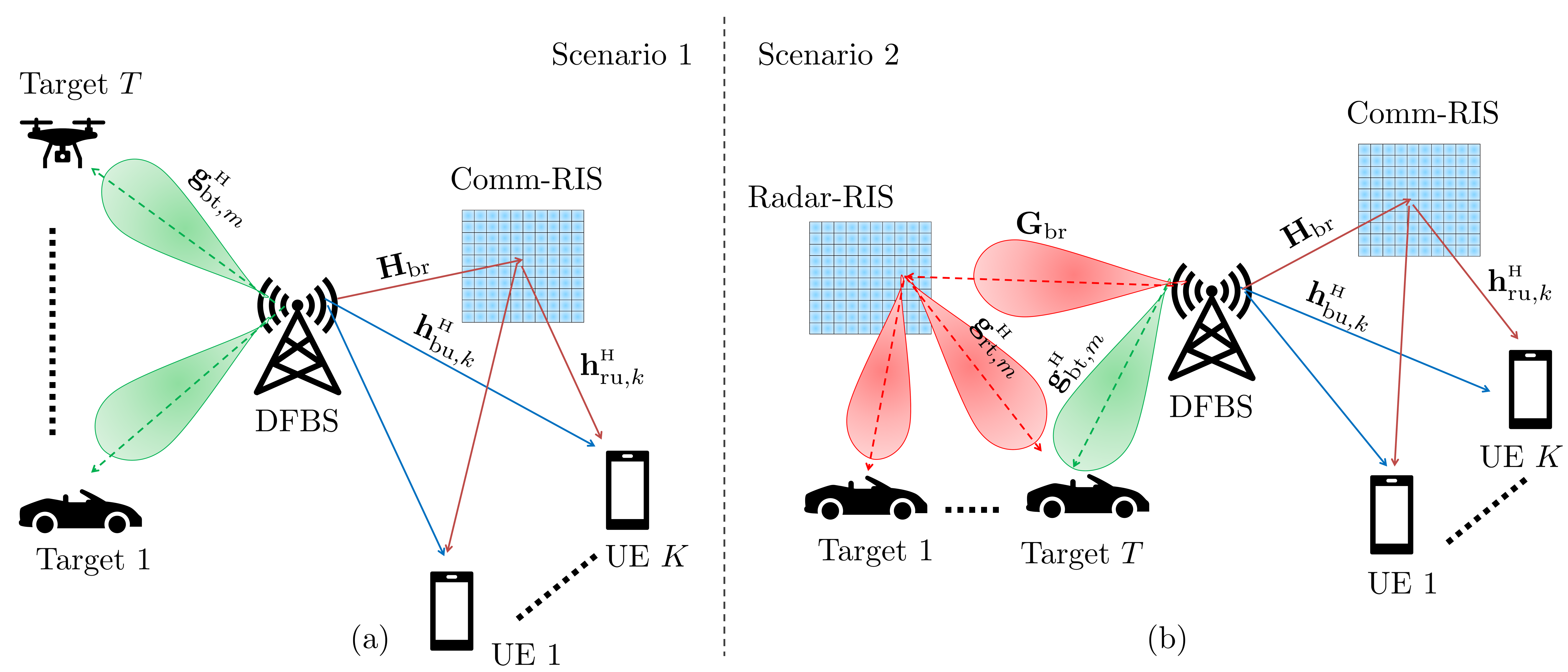}
		\caption{ System model of an \ac{ris}-assisted \acs{isac} system. (a) \ac{ris} assists only the communication functionality. (b) Dual \acp{ris} for communications and sensing.} 
		\label{fig:sys_model}
	\end{figure*}
	
 In this paper, we focus on multi-user multi-target scenarios and propose two algorithms to circumvent the limitations in the prior art, such as the possibility of \ac{isac} systems missing a few targets and enabling sensing even when the direct path to the targets is absent. To this end, we  form multiple beams towards all target directions  by using a fairness-promoting radar metric to ensure that all the targets are illuminated. We propose to use an additional dedicated \ac{ris} to enable reliable sensing even when the direct paths to some or all of the targets are blocked. Major contributions in the paper are summarized as follows. 	\\
	{\bf \ac{ris} only for assisting communications:}
	In the first setting, a single \ac{ris}, referred to as comm-\ac{ris}, is solely used to enhance the communication performance. We assume that the targets and users are well separated with the comm-\ac{ris} located closer to the \acp{ue}. We design the \ac{ris} phase shifts and transmit beamformers to form uncorrelated beams by matching it to a desired beampattern while ensuring a minimum received \ac{sinr} for the \acp{ue}. This is a non-convex optimization problem. Therefore, we design the beamformers and phase shifts in an alternating manner by designing the transmit beamformers while keeping the \ac{ris} phase shifts fixed and vice versa.  Given the beamformers, we design the comm-\ac{ris} phase shifts to maximize the \ac{sinr}, which leads to a \emph{multi-ratio fractional problem}. We solve the  multi-ratio fractional problem using an iterative procedure based on the \textit{Dinkelbach method}~\cite{dinkelbach1967non_linear_fractional}.
 Due to the specific choice of radar and communication metrics,  the proposed method also enforces fairness to all the  \acp{ue} and targets.\\
	{\bf Dual \ac{ris} for both communications and sensing:} In the second setting, we propose a dual-RIS assisted \ac{isac} system, where we use  geographically separated dedicated \acp{ris} for sensing and communication.  As before, we assume that the users and targets are spatially well separated and that a single comm-\ac{ris} located close to the \acp{ue} cannot efficiently sense the targets, hence motivating the need to use an additional \ac{ris}, referred to as radar-\ac{ris}, for sensing. Due to the fully passive nature of both \acp{ris}, we cannot form uncorrelated beams for radar sensing as before. Hence, we maximize the worst-case target illumination power while ensuring a minimum received SINR for the \acp{ue}. 
	We solve the resulting non-convex design problem using alternating optimization.  While the procedure for designing the comm-RIS phase shifts is the same as in the first setting, to design the radar-RIS phase shifts, we propose to use a method based on semi-definite relaxation followed by Gaussian randomization.  In contrast to~\cite{hua2022joint}, our scheme ensures that even the weakest target is illuminated  so that the \ac{dfbs} can reliably sense all the targets present in the scene even if the targets are not directly visible to the \ac{dfbs}.
	
	Through numerical simulations, we demonstrate the benefits of the proposed design of \ac{ris}-assisted \acs{isac} systems. The comm-\ac{ris} assisted \acs{isac} system  significantly improves the fairness \ac{sinr} of the communication \acp{ue}~(often by about $15-20$ dB when compared with~\cite{hua2022joint}  in the considered simulation setting) while suffering a moderate loss of about $1.5-2.5$ dB in the worst-case target illumination power as compared to systems without \ac{ris}~\cite{liu2020joint_transmit_beamform}. This loss is due to the fraction of the total power sent to the comm-\ac{ris}. For the dual-RIS assisted setup with dedicated \acp{ris} for sensing and communications, due to the enhanced degrees of freedom in the design due to  the radar-RIS, both the fairness \ac{sinr} and worst-case target illumination power are significantly improved~(by about $10~{\rm dB}$ and  $15-20~{\rm dB}$, respectively, in the considered simulation setting) as compared with an \acs{isac} system without RIS~\cite{liu2020joint_transmit_beamform}, especially when all the targets are not directly visible to the \ac{dfbs}.

		The remainder of the paper is organized as follows. We present the system model in Section~\ref{sec:sysmodel}. We describe the performance metrics and formulate the design problems in Section~\ref{sec:problem_formulation}. The proposed solvers are developed in Section~\ref{sec:Design1}.  Numerical simulations demonstrating the performance of the proposed algorithms are presented in Section~\ref{sec:simulations}. We conclude the paper in Section~\ref{sec:conclusion}.

	Throughout the paper, we use lowercase letters to denote scalars and boldface lowercase (uppercase) to denote vectors (matrices). We use  $(\cdot)\rT$, ${(\cdot)^{*}}$, and $(\cdot)\rH$ to denote transpose, complex conjugation, and Hermitian (i.e., complex conjugate transpose) operations, respectively and  $[\vx]_n$ or $x_n$ is the $n$-th entry of the vector~$\vx$. We define the set of $N$ dimensional vectors with unit modulus entries as $\Omega^N = \{ \va \in \mathbb{C}^N \,: \, |a_i| = 1,~i=1,\ldots,N\}$ and the set of all $M \times M$ positive semi-definite matrices are denoted by $\mathbb{S}_+^M$.

	\section{System model} \label{sec:sysmodel}

	Consider an \ac{ris}-assisted \ac{mimo} \acs{isac} system with a \ac{dfbs} communicating with $K$ single antenna users and simultaneously transmitting radar waveforms towards $T$ point targets in the far field of the \ac{dfbs} and \ac{ris}. The \ac{dfbs} has a \ac{ula} with $M$ antennas, which transmits both communication symbols and radar waveforms.   We assume that the targets and users are spatially well separated with the comm-\ac{ris}~(the radar-\ac{ris}) located closer to the \acp{ue}~(respectively, the targets) as illustrated in Fig.~\ref{fig:sys_model}. Hence, the wireless links between the comm-\ac{ris}~(the radar-\ac{ris}) and the targets~(respectively, the \acp{ue}) are very weak, and effect of the corresponding wireless channels on the system can be safely ignored. The first setting is well suited when the targets are directly visible to the \ac{dfbs} via a direct path, like for aerial surveillance or for environments with limited scattering. On the other hand, the second setting is suited for scenarios where the direct path to the targets is blocked or very weak, e.g., localization of cars or pedestrians in an urban setting, i.e., for environments with rich scattering.  
	
	\subsection{Downlink transmit signal model} \label{sec:sysmodel:dl_tx}
	Let $\vw_n = [w_1[n],\ldots,w_M[n]]\rT \in \mbC^{M}$ denote the discrete-time baseband radar waveforms at time instance~$n$.  We define the discrete-time complex baseband downlink communication symbols transmitted to the $K$ UEs as $\vd_n = [d_1[n],\ldots,d_K[n]]\rT \in \mbC^{K}$. We precode $\vd_n$ with the communication beamformer $\mC = [\vc_1,\vc_2,\ldots,\vc_K] \in \mbC^{M \times K}$ and precode $\vw_n$ using the sensing beamformer $\mS = [\vs_1,\vs_2,\ldots,\vs_M] \in \mbC^{M \times M}$. The precoded radar waveforms and communication symbols are superimposed and transmitted from the \ac{dfbs}. The complex baseband signal transmitted by the \ac{dfbs} is 
	\begin{equation} \label{eq:transmit_signal}
		\vx_n = \mC \vd_n + \mS \vw_n \, \in \mbC^{M}.
	\end{equation}
	We assume that the \acp{ue} do not cooperate with each other and hence the communication symbols transmitted towards different \acp{ue} are assumed to be uncorrelated. We also assume that the radar and communication symbols are uncorrelated with each other and have unit power, i.e., $\mR_{d} = \mbE \left[ \vd_n \vd_n\rH \right] = \mI_{M}$, $\mR_{w} = \mbE \left[ \vw_n \vw_n\rH \right] = \mI_{M}$, and  $\mbE \left[ \vd_n \vw_n\rH \right] = \boldsymbol{0}$. Thus, the transmit signal covariance matrix $\mR = \mbE \left[ \vx_n \vx_n\rH \right] \in \mathbb{S}_+^M$ is given by
	\begin{equation} \label{eq:transmit_signal_covariance}
		\begin{aligned}
			\mR = \mC\mC\rH + \mS\mS\rH &= \sum_{k=1}^{K}\vc_k \vc_k \rH + \mS\mS\rH \\
			&=\sum_{k=1}^{K}\mC_k + \mS\mS\rH,
		\end{aligned}	
	\end{equation}
	where we have introduced the rank-1 matrix $\mC_k =\vc_k\vc_k\rH$.

	\subsection{RIS model} \label{sec:sysmodel:ris}
	
	We model the \ac{ris} as a collection of discrete passive phase shifters, which can be individually controlled from the \ac{dfbs} via a low-rate control link. Specifically, we assume that the RIS comprises $N$ elements and model its spatial response as that of a \ac{ura}. The phase shifts of the RIS are collected in the vector $\boldsymbol{\omega}_t = [\omega_{t,1},\ldots,\omega_{t,N}]\rT~\in~\Omega^{N}$, where $\omega_{t,i}$ denotes the phase shift introduced by the $i$th RIS element. Here, $t \in~\{{\rm c},{\rm r}\}$ with ${\rm c}$ and ${
		\rm r}$ denoting the comm-RIS and radar-RIS, respectively. 
	
	In the first setting, the \ac{ris} empowers only the communication part of \acs{isac}, wherein the transmit signal reaches the \acp{ue} via a direct and an indirect path through the RIS and the transmit signal reaches the target only via a direct path. In the second setting, we have a dual-\ac{ris} setup with one \ac{ris} dedicated to sensing and one to communications.  In the second setting, in addition to both users and targets receiving signals via the direct path, users~(targets) also receive signal through the comm-\ac{ris}~(respectively, the radar-\ac{ris}). 
	
	\subsection{Communication channel model}  \label{sec:sysmodel:comm:channel}
	We consider a multi-user MIMO setting in which the \ac{dfbs} communicates with $K$ single antenna \acp{ue}. 	
	Let $\mH_{\rm br} \in \mbC^{N \times M}$ denote the MIMO channel matrix for the \ac{dfbs}-comm-\ac{ris} link. Let $\vh_{{\rm bu},k}\rH \in \mbC^{1 \times M}$ and $\vh_{{\rm ru},k}\rH \in \mbC^{1\times N}$ denote the multiple input single output~(MISO) channel vectors of the $k$th UE corresponding to the \ac{dfbs}-UE and \ac{ris}-UE links, respectively. The overall channel vector $\vh_k\rH$ comprising of the direct link and the additional  link via the RIS is given by
	\begin{equation} \label{h_k_def}
		\vh_k\rH = \vh_{{\rm bu},k}\rH + \vh_{{\rm ru},k}\rH{\rm diag}(\boldsymbol{\omega}_{\rm c}) \mH_{\rm br} \in \mbC^{1 \times M}.
	\end{equation}
	The signal received at the $k$th \ac{ue} is then
	\begin{align}
		y_k &= \vh_k\rH \vx_n + n_k = \vh_k\rH \left( \mC \vd_n + \mS \vw_n  \right) + n_k, \label{ue_rec_signal}
	\end{align}
	where $n_k$ is the additive Gaussian receiver noise having zero mean and variance $\sigma^2$.  
	
	\subsection{Radar channel model} \label{sec:sysmodel:radar:channel}
	Let $\theta_j$ denote the angular location of the $j$th target \ac{wrt} the \ac{dfbs}. Then the \ac{dfbs}-target channel for the $j$th target is modeled as an \ac{los} channel and is given by
	\begin{equation}\label{g_bt_definition}
		\vg_{{\rm bt},j}\rH =  \alpha_{\rm bt,j} \va\rH(\theta_j) \in \mathbb{C}^{1 \times M},
	\end{equation} 
	where $\alpha_{\rm bt,j}$ is the complex path gain and $\va(\cdot)$ is the array response vector of the \ac{dfbs}. We denote the channel from the radar-\ac{ris} to the $j$th target by $\vg\rH_{{\rm rt},j} \in \mathbb{C}^{1 \times N}$. The overall channel between the \ac{dfbs} and the $j$th target is then 
	\begin{equation} \label{radar_channel_s2}
		\vg_j\rH = \vg_{{\rm bt},j}\rH  + \vg_{{\rm rt},j}\rH {\rm diag}(\boldsymbol{\omega}_{\rm r})\mG_{\rm br} \in \mathbb{C}^{1 \times M},
	\end{equation}
	where $\mG_{\rm br} \in \mathbb{C}^{N \times M}$ is the channel between the \ac{dfbs} and radar-\ac{ris}. Whenever the radar-RIS is not used or available, we simply set $\boldsymbol{\omega}_{\rm r}$ to zero.

	\section{Problem formulation} \label{sec:problem_formulation}
	
	The performance of an ISAC system is characterized by the communication and radar sensing metrics, which are used to design the transmit beamformers and the \ac{ris} reflection patterns.
	
	\subsection{Communications metric} \label{sec:problem_formulation:comm_metric}
	
	The quality of service for the multi-user MIMO communication system is determined by the spectral efficiency or rate of the \acp{ue}. For multi-user MIMO systems, the spectral efficiency or rate of the \acp{ue} is determined by the \ac{sinr} at each user. To ensure a minimum SINR for all the \acp{ue}, we consider the worst-case SINR or the so-called fairness SINR as the communication metric.
	
	To arrive at an expression for the \ac{sinr}, let us express the received signal at the $k$th UE, i.e.,~\eqref{ue_rec_signal} as
	\begin{equation*} \label{eq:received_signal_long}
		y_k =	\vh_k\rH \left( \vc_k d_k[n]  + \sum_{j=1,j\neq k}^{K} \vc_j d_j[n]  + \sum_{m=1}^{M} \vs_m w_m[n] \right) + n_k,
	\end{equation*}
	where the first term corresponds to the intended symbol received at the $k$th UE, the second term corresponds to the interference arising from the communication symbols intended for other \acp{ue}, and the third term corresponds to the interference arising from radar waveforms. Thus, the SINR for the $k$th UE is a function of ${\boldsymbol \omega}_c, \mC,\mS$ and is given by
	
	\begin{align}
		\gamma_k({\boldsymbol \omega}_c,\mC,\mS) &= \frac{ \vert \vh_k\rH \vc_k  \vert^2  }{ \sum_{j=1,j\neq k}^{K} \vert \vh_k\rH \vc_j \vert^2 + \sum_{m=1}^{M}\vert \vh_k\rH \vs_m  \vert^2 + \sigma^2} \nonumber\\
		&=\frac{ \vh_k\rH  \vc_k \vc_k \rH \vh_k }{\vh_k\rH \left( \sum_{j=1,j\neq k}^{K} \vc_j \vc_j\rH + \mS\mS\rH   \right)\vh_k + \sigma^2 } \nonumber\\
		&= \frac{ \vh_k \rH \mC_k \vh_k }{\vh_k \rH \left( \mR - \mC_k \right)\vh_k + \sigma^2} \label{sinr_k}
	\end{align}
	
	with the worst-case SINR 
	\begin{equation} \label{commun_metric}
		\gamma_{\rm min} = \underset{k}{\min}\quad \gamma_k({\boldsymbol \omega}_c,\mC,\mS).
	\end{equation}
	
	\subsection{Radar sensing metrics}
	Next, we describe sensing performance metrics that we use for the two scenarios: when we do not use radar-\ac{ris} and when we use radar-\ac{ris}. 

	\subsubsection{Beampattern matching and cross-correlation error} \label{sec:problem_formulation:radar_metric}
	A MIMO radar system detects and tracks targets by transmitting signals in certain desired directions. This is achieved by designing multiple beams  that match a desired beampattern while keeping the correlation between different beams as small as possible~\cite{stoica2007on_probing_signal,liu2020joint_transmit_beamform}.
	
	To ensure that sufficient power reaches different target locations, we need to choose a desired pattern $d(\theta)$ that has main beams pointing in the direction of different targets. Furthermore, the comm-RIS is useful only if sufficient power reaches the comm-RIS in the first place. Thus, the desired pattern should also form a beam towards the direction of the comm-RIS from the \ac{dfbs}, denoted by $\zeta_{\rm r}$. Hence, the desired pattern is chosen to be a superposition of multiple rectangular beams of width $\epsilon$ degrees with centers around $\{\theta_k\}_{k=1}^{T}$ and $\zeta_{\rm r}$ as
	\begin{equation} \label{d_theta_design}
		d({\theta}) = \begin{cases}
			1, & \text{if} \quad {\theta} \in \left[ \theta_k - \epsilon, \theta_k + \epsilon \right], \\
			1, & \text{if} \quad {\theta} \in \left[ \zeta_{\rm r} - \epsilon, \zeta_{\rm r} + \epsilon \right], \\ 
			0, & \text{elsewhere}.
		\end{cases}
	\end{equation}
	The power radiated from the \ac{dfbs} towards the direction $\theta$ is given by
	\begin{equation}
		J(\theta) = \mbE \left[ \vert \va\rH (\theta) \vx_n \vert^2 \right] = \va \rH (\theta) \mR \va(\theta).
	\end{equation}

	For a desired beampattern $d(\theta)$, the beampattern mismatch error evaluated over a discrete grid of $L$ angles $\{\tilde{\theta}_i\}_{i=1}^{L}$ is a function of $\mR$ and is given by
	\begin{equation}
		L_1(\mR,\tau) = \frac{1}{L}\sum_{\ell=1}^{L} \vert J(\tilde{\theta}_{\ell}) - \tau d(\tilde{\theta}_{\ell})  \vert^2,
	\end{equation}
	where $\tau$ is the unknown autoscale parameter. The
	cross-correlation of the signals reflected back by the targets at directions $\tilde{\theta}_i$ and $\tilde{\theta}_j$ is given by
	\begin{equation*}
		J_c(\mR,\tilde{\theta}_i,\tilde{\theta_j}) = \mbE\left[ (\va\rH(\tilde{\theta}_i)\vx_n) (\vx_n\rH  \va(\tilde{\theta}_j))  \right] = \va\rH(\tilde{\theta}_i)\mR \va(\tilde{\theta}_j),
	\end{equation*}
	with the average squared cross-correlation being
	\begin{equation}
		L_2(\mR) = \frac{2}{T^2 - T}\sum_{i=1}^{T-1} \sum_{j=i+1}^{T} \vert 	J_c(\mR,\tilde{\theta}_i,\tilde{\theta_j})  \vert^2.
	\end{equation}
	We design radar beams that minimize a weighted sum of beampattern mismatch error and the average squared cross-correlation~\cite{stoica2007on_probing_signal}
	\begin{equation} \label{eq:radar_cost1}
		L(\mR,\tau) = w_b L_1(\mR,\tau) + w_c L_2(\mR),
	\end{equation}
	where $w_b$ and $w_c$ are the known weights that determine the relative importance of the two terms.  
	
	\subsubsection{Worst-case target illumination power}
	Whenever the direct paths between the \ac{dfbs} and the targets are weak or blocked, forming multiple beams at the \ac{dfbs} towards different targets by adopting a beampattern mismatch criterion will not be useful as the target illumination power will be very small. To enable sensing in such scenarios, radar-\ac{ris} can be used. However, due to the fully passive nature of the \ac{ris}, obtaining cross-correlation optimal beams directed towards targets is not possible. Therefore, we adopt an alternative metric and propose to maximize the worst-case target illumination power to design the sensing beamformers. 
	
	The power of the signal received at the $m$th target corresponding to the transmitted waveform $\vx_n$ is $\mbE\left[ \vert \vg_m\rH\vx_n \vert^2 \right]$. The worst-case target illumination power is given by
	\begin{align}
		Q(\mR,\boldsymbol{\omega}_{\rm r}) &= \min_m \quad \mbE\left[ \vert \vg_m\rH\vx_n \vert^2 \right] \nonumber \\
		&=  \min_m \quad  \vg_m \rH \mR \vg_m  
		=  \quad \min_m  \quad {\rm Tr} \left(\mR \mD_m \right), \label{eq:radar_cost2}
	\end{align}
	where $\mD_m = \vg_m \vg_m\rH$ and $Q$ depends on $\boldsymbol{\omega}_{\rm r}$ through $\{\vg_m, m=1,\ldots,M\}$ [c.f.~\eqref{radar_channel_s2}].

	\subsection{ Design problems  }
	We now introduce the problems of designing the transmit beamformers at the \ac{dfbs} and the \ac{ris} phase shifts for the two settings.
	For the setting [cf. Fig.~\ref{fig:sys_model}(a)] with the comm-RIS assisting \ac{dfbs} in  communicating with multiple users, we design the transmit beamformers ($\mC,\mS$) and the comm-\ac{ris} phase shifts $\boldsymbol{\omega}_{\rm c}$ by minimizing the sensing cost function \eqref{eq:radar_cost1} while ensuring a minimum SINR for all users. For the setting [cf. Fig.~\ref{fig:sys_model}(b)], we propose to design the transmit beamformers ($\mC,\mS$) and the radar- and comm-\ac{ris} phase shifts $\boldsymbol{\omega}_{\rm c}$ and $\boldsymbol{\omega}_{\rm r}$,  to maximize the worst-case target illumination power $Q(\mR,\boldsymbol{\omega}_{
			\rm r})$ [cf.~\eqref{eq:radar_cost2}] while ensuring a minimum SINR for all the users. We  mathematically formulate the design problem in the first setting as
		\begin{subequations}   
			\label{problem_formulation}
			\begin{align} 
				\quad \underset{\mS,\mC,\tau,\boldsymbol{\omega}_{\rm c}}{\text{minimize}} & \quad L(\mR,\tau)  \nonumber \\
				(\mathcal{P}1):  \,\, \text{subject to} & \quad \mR = \mC\mC\rH + \mS\mS\rH \,\in\, \mathbb{S}_+^M \label{problem_formulation_psd}\\
				& \quad [\mR]_{i,i} = P_t/M, \,\, i=1,\ldots,M   \label{problem_formulation_pow}\\
				& \quad \gamma_k(\boldsymbol{\omega}_{\rm c},\mC,\mS) \geq \Gamma, \,\, k=1,\ldots,K  \label{problem_formulation_sinr}\\
				& \quad {\boldsymbol \omega}_{\rm c} \in \Omega^N,  \label{problem_ris_constraint} 
			\end{align}
		\end{subequations}
		and the second setting  as
		\begin{subequations} \label{problem_formulation_s2}
			\begin{align} 
				\quad \underset{\mC,\mS,\boldsymbol{\omega}_{\rm c},\boldsymbol{\omega}_{
						\rm r}}{\text{maximize}} & \quad Q(\mR,\boldsymbol{\omega}_{
					\rm r})  \nonumber\\
				(\mathcal{P}2):  \,\,	\text{subject to} & \quad \mR = \mC\mC\rH + \mS\mS\rH \,\in\, \mathbb{S}_+^M \label{radar_ris_sp1_psd} \\
				& \quad [\mR]_{i,i} = P_t/M, \,\, i=1,\ldots,M \label{radar_ris_sp1_pow}\\
				& \quad \gamma_k(\boldsymbol{\omega}_{\rm c},\mC,\mS) \geq \Gamma, \,\, k=1,\ldots,K  \label{radar_ris_sp1_sinr}\\
				& \quad {\boldsymbol \omega}_{\rm r} \in \Omega^N, \, {\boldsymbol \omega}_{\rm r} \in \Omega^N,  \label{ris_sp1}
			\end{align}
		\end{subequations}
		where $P_t$ is the total transmit power with \eqref{problem_formulation_pow} and~\eqref{radar_ris_sp1_pow} denoting the power constraint per antenna element, $\Gamma$ is the SINR requirement of each user, and recall that $\tau$ is the autoscale parameter. The constraints~\eqref{problem_ris_constraint} and \eqref{ris_sp1} are due to the fact that the \ac{ris} is fully passive and can only reflect (i.e., phase shift) the incident signal to different directions. The optimization problem~\eqref{problem_formulation} is non-convex because of the quadratic equality in~\eqref{problem_formulation_psd}, fractional term in~\eqref{problem_formulation_sinr}, and unit-modulus constraint in~\eqref{problem_ris_constraint}. 
		
		In contrast to $(\cP1)$, we use a different objective function in $(\cP2)$. In addition to the design of the transmit beamformers and comm-\ac{ris} phase shifts, we also design the radar-\ac{ris} phase shifts in ($\cP2$). Optimization problem $(\mathcal{P}2)$ is also non-convex because of the quadratic equality~\eqref{radar_ris_sp1_psd}, fractional term in the constraint~\eqref{radar_ris_sp1_sinr}, and unit-modulus constraints~\eqref{ris_sp1}.

	\section{  Proposed solvers  }  \label{sec:Design1}
	In this section, we develop the proposed solvers for $(\cP1)$ and $(\cP2)$. 

	\subsection{ Proposed solver for $(\mathcal{P}1)$} 
	
	Problem $(\mathcal{P}1)$ to design the transmit beamformers and comm-\ac{ris} phase shifts is not convex in the variables. Therefore, we propose to solve the optimization problem by alternatingly optimizing the beamformers and the autoscale parameter (i.e., $\mC,\mS,\tau$) while keeping the phase shifts (i.e., ${\boldsymbol \omega}_{\rm c}$) fixed, and vice versa. Specifically, in the first subproblem, we fix the comm-\ac{ris} phase shifts and design the beamformers to minimize the beampattern mismatch error while ensuring a minimum SINR for the users. Next, keeping the beamformers fixed, we optimize the comm-\ac{ris} phase shifts to maximize the worst-case \ac{sinr} for all the \acp{ue}. 

	We repeat the aforementioned steps till convergence.

	\subsubsection{Updating $\mC$, $\mS$ and $\tau$, given $\boldsymbol{\omega}_{\rm c}$} \label{alg1_cs_update}
	Let us now consider the subproblem of designing the beamformers $\mC$ and $\mS$ by fixing comm-\ac{ris} phase shifts $\boldsymbol{\omega}_{\rm c}$ for an appropriately selected SINR threshold $\Gamma$. That is, we solve~\cite{liu2020joint_transmit_beamform}:
	\begin{equation} \label{subproblem1_initial}
		\begin{aligned} 
			\quad	\underset{\mS,\mC,\tau}{\text{minimize}} & \quad L(\mR,\tau) \\
			\text{subject to} & \quad \mR = \mC\mC\rH + \mS\mS\rH \,\in\, \mathbb{S}_+^M   \\
			& \quad [\mR]_{i,i} = P_t/M, \,\, i=1,\ldots,M  \\
			& \quad \gamma_k (\mR) \geq \Gamma, \,\, k=1,\ldots,K,  
		\end{aligned}
	\end{equation}
	which is still a non-convex optimization problem. To solve~\eqref{subproblem1_initial}, we use the procedure from~\cite{liu2020joint_transmit_beamform}, which is provided here for self containment. From~\eqref{eq:transmit_signal_covariance} and~\eqref{sinr_k}, the \ac{sinr} constraints can be expressed as
	\begin{equation}
		\left( 1 + \Gamma^{-1}\right) \vh_k \rH \mC_k \vh_k \geq \vh_k \rH \mR \vh_k + \sigma^2 \label{sp1_sinr_def2}
	\end{equation}
	for $k=1,\ldots,K$. Thus the subproblem~\eqref{subproblem1_initial} can be equivalently written as
	
	\begin{align} 
		\underset{\mR,\mC_1, \ldots, \mC_K, \mS,\tau}{\text{minimize}} & \quad L(\mR,\tau) \nonumber \\
		\text{subject to} & \quad \mR = \sum_{k=1}^{K} \mC_k + \mS\mS\rH,  \mR - \sum_{k=1}^{K}{\mC}_k \,\in\, \mathbb{S}_+^M,  \nonumber \\
		& \quad  \mR \in \mathbb{S}_+^M, \nonumber \\
		& \quad \left( 1 + \Gamma^{-1}\right) \vh_k \rH \mC_k \vh_k \geq \vh_k \rH \mR \vh_k + \sigma^2, \nonumber \\
		& \quad {\rm rank}(\mC_k) = 1, \quad k=1,\ldots, K \nonumber \\
		& \quad [\mR]_{i,i} = P_t/M, \quad i=1,\ldots,M. \label{sp2} 
		\end{align}	
	By dropping the quadratic equality constraint and rank constraints, we obtain the following relaxed convex optimization problem
	\begin{align} 
		\underset{\mR,\mC_1, \ldots, \mC_K, \tau}{\text{minimize}} & \quad L(\mR,\tau) \nonumber \\
		\text{subject to} \quad & \quad \mR \in \mathbb{S}_+^M ,\mR - \sum_{k=1}^{K}{\mC}_k \in \mathbb{S}_+^M   \nonumber \\
		& \quad [\mR]_{i,i} = P_t/M, \quad i=1,2,\ldots,M \nonumber \\
		& \quad \left( 1 + \Gamma^{-1}\right) \vh_k \rH \mC_k \vh_k \geq \vh_k \rH \mR \vh_k + \sigma^2, \nonumber \\
		& \quad \mC_k \in \mathbb{S}_+^M , \quad k=1,\ldots, K, \label{sp3} 
	\end{align}	
	which is a semi-definite quadratic program~(SQP) that can be efficiently solved using off-the-shelf solvers.  
	
	Let us call the solution of~\eqref{sp3} as $\hat{\mR}, \tilde{\mC}_1,\ldots,\tilde{\mC}_K$. Now we compute the beamformers $\hat{\mC} = [\hat{\vc}_1,\ldots,\hat{\vc}_K] \in \mbC^{M \times K}$ as
	\begin{equation} \label{sp3_c_construct}
		\hat{\vc}_k = \frac{\tilde{\mC}_k\vh_k}{\sqrt{ \vh_k \rH \tilde{\mC}_k \vh_k }}, \quad \hat{\mC}_k = \hat{\vc}_k \hat{\vc}_k\rH, \quad k=1,\ldots,K
	\end{equation} 
	and $\hat{\mS} = [\hat{\vs}_1,\ldots,\hat{\vs}_M] \in \mbC^{M \times M}$ using the Cholesky decomposition as
	\begin{equation} \label{sp3_s_construct}
		\hat{\mR} - \sum_{k=1}^{K} \hat{\mC}_k := \hat{\mS} \hat{\mS}\rH. 
	\end{equation}
	By construction, $\hat{\mC}_k$ satisfies the \ac{sinr} constraints as $\vh_k \rH \hat{\mC}_k \vh_k = \vh_k \rH \tilde{\mC}_k \vh_k$. Since $\hat{\mR} - \sum_{k=1}^{K}\tilde{\mC}_k \,\in\, \mathbb{S}_+^M $ and
	\[
	\vz\rH(\tilde{\mC}_k - \hat{\mC}_k)\vz =  \vz\rH \tilde{\mC}_k \vz - (\vh_k\rH \tilde{\mC}_k \vh_k)^{-1} \vert \vz\rH \tilde{\mC}_k \vh_k \vert^2 \geq 0
	\] 
	for any $\vz$ due to the Cauchy-Schwartz inequality, we have $\hat{\mR} - \sum_{k=1}^{K}\tilde{\mC}_k +  \sum_{k=1}^{K} (\tilde{\mC}_k - \hat{\mC}_k) = \hat{\mR} - \sum_{k=1}^{K} \hat{\mC}_k \,\in\, \mathbb{S}_+^M .$ Thus, $\hat{\mC}_k, k=1,2,\ldots, K$ also form a feasible solution to~\eqref{sp3}.  Further, the constructed solution does not change the objective value, and thus will yield the smallest objective value for~\eqref{sp3}.

	\subsubsection{Updating $\boldsymbol{\omega}_{\rm c}$,  given $\mC$, $\mS$, and $
		\tau$} \label{design1:step2}
	The subproblem of finding the comm-\ac{ris} phase shifts, given $\mC$, $\mS$, and $
	\tau$, simplifies to the feasibility problem
	\begin{align}
		&{\text{find}} \quad \quad  \quad \,\,\, \boldsymbol{\omega}_{\rm c}  \nonumber \\
		& \text{subject to} \quad    \gamma_k(\boldsymbol{\omega}_{\rm c}) \geq \Gamma, \,\, k=1,\ldots,K  \nonumber\\
		& \quad \quad  \quad \quad \quad \, {\boldsymbol \omega}_{\rm c} \in \Omega^N. \label{eq:feasiblity_problem_w}
	\end{align}
	However, the fixed choice of $\boldsymbol{\omega}_{\rm c}$ in \eqref{subproblem1_initial} is already a solution to the above problem. Since different choices of $\boldsymbol{\omega}_{\rm c}$ lead to different achievable \acp{sinr} at the \acp{ue}, i.e., better $\Gamma$, we can alternatively design $\boldsymbol{\omega}_{\rm c}$ to improve the achievable \ac{sinr} at the \acp{ue} by solving
	\begin{align} \label{subproblem2}
		\Gamma_1 = \quad	\underset{\boldsymbol{\omega}_{\rm c} \in \Omega^N}{\text{maximize}} \quad \underset{1\leq k \leq K}{\min}& \quad \gamma_k (\boldsymbol{\omega}_{\rm c}) 
	\end{align}
	where $\Gamma_1 \geq \Gamma$. Due to the unit modulus constraint and the fractional nature of the cost function,~\eqref{subproblem2} is a non-convex optimization problem in $\boldsymbol{\omega}_{\rm c}$ and its exact solution is difficult to compute. 
	
	To solve~\eqref{subproblem2}, we first express it as a generalized linear fractional program with multiple ratios by expressing 
	the \ac{sinr} explicitly in terms of $\boldsymbol{\omega}_{\rm c}$. Define the vectors 
	$\va_{k,m} = [ {\rm diag}(\vh_{{\rm ru},k}\rH)\mH_{\rm br} \vc_m , \vh_{{\rm bu},k} \rH \vc_m]\rT$ 
	and 
	$\vb_{k,m} = [{\rm diag}(\vh_{{\rm ru},k}\rH)\mH_{\rm br} \vs_m ,  \vh_{{\rm bu},k}\rH \vs_m]\rT$ of length $N+1$. 
	Let us also define the square matrices $\mA_k = \va_{k,k} \va_{k,k}\rH$ and $\mB_k  = \sum_{j=1,j\neq k }^{K} \va_{k,j}\va_{k,j}\rH + \sum_{m=1}^{M} \vb_{k,m} \vb_{k,m}\rH$ of size $N+1$. From~\eqref{h_k_def}, we can then express \ac{sinr} in \eqref{sinr_k} as
	\begin{align}
		\gamma_k(\vw) = \frac{  \vw\rH \mA_k \vw}{\vw\rH \mB_k \vw + \sigma^2} =   \frac{ {\rm Tr}\left(  \mA_k \vw\vw\rH \right) }{  {\rm Tr} \left(  \mB_k \vw  \vw\rH \right) + \sigma^2} 
		\end{align}
	where $\vw\rH = [\boldsymbol{\omega}_{\rm c}\rT ,1 ]$. By letting $\mW  = \vw \vw\rH$, we rewrite~\eqref{subproblem2} as
	\begin{align}
		\Gamma_2  = \quad	&\underset{\mW \in \mathbb{S}_+^{N+1} }{\text{maximize}} \quad \underset{1 \leq k \leq K}{\min} \quad \frac{ {\rm Tr}\left(  \mA_k \mW \right) }{  {\rm Tr} \left(  \mB_k \mW \right) + \sigma^2} \nonumber\\
		&\text{subject to}  \quad 
		[\mW]_{ii} = 1, \quad i=1,\ldots,N+1, \label{subproblem2_mod3}
	\end{align}     
	where we have introduced a less restrictive and convex semi-definite constraint on $\mW$ instead of the rank one constraint. The problem in \eqref{subproblem2_mod3} is a generalized linear fractional program with a convex constraint set. Next, we develop an iterative procedure to solve ~\eqref{subproblem2_mod3} using the generalized Dinkelbach-type method~\cite{Crouzeix1991algorithms_gfp}, which is guaranteed to converge to the solution of~\eqref{subproblem2_mod3}.

		Let us denote the iterate at step $t$ as $\mW^{(t)}$. We first compute an estimate of the worst-case \ac{sinr} as
		\begin{align}
			\label{eq:dinkelbach_1}
			\lambda^{(t)} = \underset{1\leq k \leq K}{ \min } \quad \frac{ {\rm Tr}\left(  \mA_k \mW^{(t-1)} \right) }{  {\rm Tr} \left(  \mB_k \mW^{(t-1)} \right) + \sigma^2}.
		\end{align}
		Next, we update $\mW^{(t)}$ as
		\begin{align}
			\label{eq:dinkelbach_2}
			\mW^{(t)}	=\,\,	&\underset{\mW \in \mathbb{S}_+^{N+1}}{\text{argmax}} \, \underset{1\leq k \leq K}{\min} \,\, {{\rm Tr}\left(  \mA_k \mW \right) } - \lambda^{(t)}{ \left(  {\rm Tr} \left(  \mB_k \mW \right) + \sigma^2 \right)} \nonumber\\
			&\text{subject to}  \quad \quad
			[\mW]_{ii} = 1, \, i=1,\ldots,N+1, %\label{subproblem2_mod3}
		\end{align}    
		where $\lambda^{(t)} \geq \Gamma_1$ and
		\[
		\min_k\left\{ {\rm Tr}(  \mA_k \mW^{(t)})  - \lambda^{(t)} \left( {\rm Tr}(  \mB_k \mW^{(t)}) + \sigma^2 \right)\right\} = 0
		\] at optimality. This condition can be used to stop the iterations (see Fig.~\ref{fig:a1_p2_converge}). In practice, we may update $\lambda^{(t)}$ and $\mW^{(t)}$ till $\|\mW^{(t)} - \mW^{(t-1)}\| \leq \texttt{Tol}$, where $\texttt{Tol}$ is the tolerance value.

	Let us denote the solution from this iterative procedure as $\tilde{\mW}$. To recover a rank-1 solution from $\tilde{\mW}$, we use Gaussian randomization~\cite{luo2010sdr}. That is, we generate $N_{\rm rand}$ realization of complex Gaussian random vectors ${
		\tilde{\vw}_i,i=1,\cdots, N_{
			\rm rand}}$ with zero mean and covariance matrix $\tilde{\mW}$. We then normalize $\tilde{\vw}_i$ to obtain unit modulus vectors $\hat{\vw}_i$ with entries 
	\[
	[\hat{\vw}_i]_n=\begin{cases}
		[\tilde{\vw}_i]_m/|[\tilde{\vw}_i]_m|, & \text{for }\, m=1,2,\ldots, N,\\
		1, & \text{otherwise.}
	\end{cases}
	\]
	We choose the realization that results in the largest worst-case \ac{sinr} as 
	\begin{align}
		i^\star = \underset{1 \leq i \leq N_{\rm rand}}{\argmax}\quad \underset{1\leq k \leq K}{\min} \quad \gamma_k(\hat{\vw}_i),
		\label{eq:omega_c_update}
	\end{align}
	and the optimal comm-\ac{ris} phase shift as $[\hat{\boldsymbol \omega}_{\rm c}]_n = [\hat{\vw}_{i^\star}^*]_n$ for $n=1,\ldots,N$. The actual worst-case SINR corresponding to the rank-1 solution obtained from this Gaussian randomization $\Gamma_3 = \min_k \{\gamma_k(\hat{\vw}_{i^\star}), k=1,\cdots,K\}$ will be in the interval $0$ and $\Gamma_1$.

	\begin{figure}[t]
		\centering
		\includegraphics[width=\columnwidth, trim=1 1 1 1, clip]{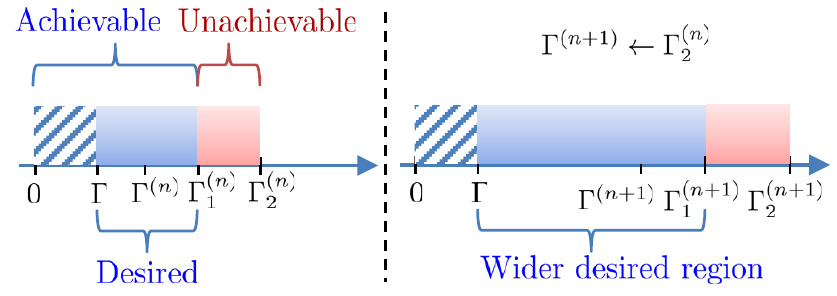}
		\caption{ Illustration of choosing $\Gamma$. The horizontal axis denotes the fairness SINR. Increasing the target SINR from $\Gamma^{(n)}$ to $\Gamma^{(n+1)} = \Gamma_2^{(n)}$ leads to an wider desired region width resulting in a higher probability of Gaussian randomization yielding a solution in the desired region.}
		\label{fig:illust_gamma}
	\end{figure}
	
	\subsubsection{Practical considerations for choosing $\Gamma$}
	\label{sec:gamma_update}
	
	Let us denote $\Gamma_1$, $\Gamma_2$ and $\Gamma_3$ at the $n$th iteration as $\Gamma_1^{(n)}$, $\Gamma_2^{(n)}$, and $\Gamma_3^{(n)}$, respectively. Since we have more degrees of freedom in choosing $\mW$ in the relaxed problem~\eqref{subproblem2_mod3} than with a unit modulus (or rank) constraint as in~\eqref{subproblem2}, we have $\Gamma_2^{(n)} \geq \Gamma_1^{(n)} \geq \Gamma$, in general. In other words, the achievable \ac{sinr} is much higher if we ignore the passive nature of the \ac{ris}, but is not practically realizable.  At each iteration, we can only evaluate $\Gamma_2^{(n)}$ and $\Gamma_3^{(n)}$, but we cannot compute $\Gamma_1^{(n)}$.
	
	To design the beamformers and the comm-\ac{ris} phase shifts, starting from an initial comm-\ac{ris} phase shift, we update $\mC$ and $\mS$ using \eqref{sp3_c_construct} and \eqref{sp3_s_construct}, respectively. Then, we update ${\boldsymbol \omega}_c$ as in \eqref{eq:omega_c_update}. Specifically, in the $n$th iteration, we solve \eqref{sp3} with the desired SINR $\Gamma$ set to $\Gamma^{(n)}$. We repeat the alternating optimization until  $\Gamma_3^{(n)} \geq \Gamma$, where recall that $\Gamma_3^{(n)} \in [0, \Gamma_1^{(n)}]$ is the actual SINR that can be achieved after Gaussian randomization. 
	We refer to the interval $[0, \Gamma_1^{(n)}]$ as the \textit{achievable region}. The interval with values of \ac{sinr} that are both achievable and desired, i.e., the interval between $\Gamma$ and $\Gamma_{1}^{(n)}$ is referred to as the \textit{desired region}. Whenever the width of the desired region is small, it is less likely that the Gaussian randomization procedure (described in the previous subsection) produces realization in the desired region. Therefore, it is possible to have $\Gamma_3^{(n)} < \Gamma$, thereby violating the communication \ac{sinr} requirements. A narrow desired region, in other words, indicates that the \ac{ris} phase shifts alone cannot significantly improve the \ac{sinr} of \acp{ue}.  At the same time, a higher \ac{sinr} is often achievable by designing beamformers since we have higher flexibility (i.e., without unit modulus constraints). To circumvent this issue, in the $(n+1)$th iteration, we design the beamformers so as to achieve an \ac{sinr} higher than $\Gamma^{(n)}$ by setting $\Gamma^{(n+1)} = \Gamma_{2}^{(n)}$, thereby increasing the width of the desired region. With a widened desired region, it is more likely that Gaussian randomization produces realization with $\Gamma_3^{(n+1)} \geq \Gamma$ as desired. This is illustrated in Fig.~\ref{fig:illust_gamma}. The proposed procedure for solving $(\mathcal{P}1)$ is summarized as Algorithm~\ref{Alg_1}.

	\subsubsection{Computational complexity}
	For an $\epsilon$-accurate solution, the worst-case complexity for computing the beamformers, given the \ac{ris} phase shifts is that of an SDP with $(K+1)$ constraints and is of the order of about $\cO\left(K^{6.5}M^{6.5}\log\left(1/\epsilon\right) \right)$~\cite{liu2020joint_transmit_beamform}. 
	For the subproblem of computing the \ac{ris} phase shifts, given the beamformers, we solve an SDP  for $I_{\rm in}$ Dinkelbach-like iterations in the inner loop. Thus, the complexity involved in obtaining the \ac{ris} phase shifts given the beamformers is of the order of about $\cO \left( I_{\rm in} N^{6.5} \log \left(1/\epsilon \right) \right)$. Suppose $I_{\rm out}$ be the number of iterations in the outer loop of the proposed alternating optimization algorithm. Then, the overall complexity is about $\cO \left( I_{\rm out} K^{6.5}M^{6.5}\log\left(1/\epsilon\right)  + I_{\rm out}I_{\rm in} N^{6.5} \log \left(1/\epsilon \right)  \right)$.

	\begin{algorithm}[!t] 
		\caption{Solver for $(\mathcal{P}1)$}\label{Alg_1}
		{\bf Initialization:} $\Gamma^{(n)} = \Gamma$, $\boldsymbol{\omega}_{\rm c}^{(n)} = \boldsymbol{\omega}_{\rm c}^{(0)}$
		\begin{algorithmic}[1]
			\For{$n=1, 2, \cdots,\texttt{MaxIter}$}
			\State Solve~\eqref{sp3} with SINR constraint set to $\Gamma^{(n)}$ to update $\hat{\mR}$ and $\{\tilde{\mC}_1,\tilde{\mC}_2, \cdots,\tilde{\mC}_K\}$. 
			\State Update $\{\hat{\mC}_1,\hat{\mC}_2, \cdots,\hat{\mC}_K\}$ using~\eqref{sp3_c_construct}. 
			\State Update $\hat{\mS}$ using~\eqref{sp3_s_construct}.
			\State Update $\boldsymbol{\omega}_{
				\rm c}$ as in~\eqref{eq:omega_c_update} and compute $\Gamma_3^{(n)}$. 
			\If{$\Gamma_3^{(n)} \geq \Gamma $} \texttt{break} \EndIf
			\State Set $\Gamma^{(n)} = \Gamma_2^{(n)} $.
			\EndFor	
		\end{algorithmic}
	\end{algorithm}

	\subsection{Proposed solver for $(\mathcal{P}2)$} \label{sec:dual_ris:precoder}
	
	We next  develop a solver to design the transmit beamformers and the phase shifts of both the radar-\ac{ris} and comm-\ac{ris}, i.e., we present a solver for Problem~$(\mathcal{P}2)$.  As before, we adopt an alternating optimization based procedure to solve the non-convex optimization problem ($\cP2$), wherein we design the beamformers while keeping the phase shifts fixed and vice versa.

	\subsubsection{Updating $\mC$ and $\mS$, given $\boldsymbol{\omega}_{\rm c}$ and $\boldsymbol{\omega}_{\rm r}$} \label{sec:P2subprob1}
	
	Given $\boldsymbol{\omega}_{\rm c}$ and $\boldsymbol{\omega}_{\rm r}$, $(\mathcal{P}2)$ simplifies to a problem with exactly the same constraints as~\eqref{sp3}, but with a different objective function, i.e., we have 
	\begin{align} 
		\underset{\mR,\mC_1, \ldots, \mC_K}{\text{maximize}} & \quad \min_m \quad {\rm Tr}\left(\mR \mD_m \right) \nonumber \\
		\text{subject to} \quad & \quad \mR \in \, \mathbb{S}_+^M ,\mR - \sum_{k=1}^{K}{\mC}_k \, \in \, \mathbb{S}_+^M  \nonumber \\
		& \quad [\mR]_{i,i} = P_t/M, \quad i=1,2,\ldots,M \nonumber \\
		& \quad \left( 1 + \Gamma^{-1}\right) \vh_k \rH \mC_k \vh_k \geq \vh_k \rH \mR \vh_k + \sigma^2, \nonumber \\
		& \quad \mC_k  \in \, \mathbb{S}_+^M , \quad k=1,\ldots, K, \label{rris_tx_pre1b_sinr} 
	\end{align}	
	where we have replaced the \ac{sinr} constraint in $(\mathcal{P}2)$ with an equivalent inequality in~\eqref{sp1_sinr_def2} and dropped the rank one constraint on $\mC_k$ as we did from \eqref{sp2} to \eqref{sp3}. A rank-1 solution $\mC_k$ and $\mS$ is obtained using~\eqref{sp3_c_construct} and \eqref{sp3_s_construct}, respectively, after solving the above convex program. The constructed solution will yield the largest objective value for \eqref{rris_tx_pre1b_sinr}.

	For a fixed $\mC$ and $\mS$, the subproblems of finding $\boldsymbol{\omega}_{\rm c}$ and $\boldsymbol{\omega}_{\rm r}$ are independent of each other and can be solved separately. 
	
	\subsubsection{Updating $\boldsymbol{\omega}_{\rm c}$, given $\mC$ and $\mS$} 
	\label{sec:dual_ris:cris}
	The procedure for updating the comm-\ac{ris} phase shifts, given the beamformers $\mC$ and $\mS$ is exactly the same as the feasibility problem~\eqref{eq:feasiblity_problem_w} for which we reuse the procedure described in Sections~\ref{design1:step2} and~\ref{sec:gamma_update}.

	\subsubsection{Updating $\boldsymbol{\omega}_{\rm r}$, given $\mC$ and $\mS$}
	\label{sec:dual_ris:rris}
	
	We now discuss the design of radar-\ac{ris} phase shifts, where we maximize the worst-case target illumination power $Q(\mR,\boldsymbol{\omega}_{\rm r})$, given the beamformers ($\mC$, $\mS$).
	
	From~\eqref{radar_channel_s2}, we can express \eqref{eq:radar_cost2} explicitly in terms of $\boldsymbol{\omega}_{\rm r}$ as
	\[
	Q(\mR,\boldsymbol{\omega}_{\rm r}) = \min_m \quad {\vu}\rH \mQ_m \vu = \min_m \quad {\rm Tr}\left(\mQ_m \mU  \right)
	\]
	where $\mU = \vu\vu\rH$ and $\vu\rH = [1, \boldsymbol{\omega}_{\rm r}\rT]$ and 
	\[
	\mQ_m =  \begin{bmatrix} \vg_{{\rm bt},m}\rH \\ {\rm diag}\left( \vg_{{\rm rt},m}^{*} \right)\mG_{\rm br}  \end{bmatrix}\mR \begin{bmatrix} \vg_{{\rm bt},m} & \mG_{\rm br}\rH {\rm diag}\left( \vg_{{\rm rt},m}\right) \end{bmatrix}.
	\]
	To obtain the optimal phase shifts, we solve the following optimization problem:
	\begin{align}
		\quad \underset{\mU \in  \mathbb{S}_+^{N+1}}{\max} & \quad \min_m \quad {\rm Tr}\left(\mQ_m \mU  \right) \label{eq:update:radar:ris} \\
		\text{subject to} &  \quad [\mU]_{i,i} = 1 \quad i=1,\ldots,N+1,	\nonumber
	\end{align}
	where we dropped the non-convex rank constraint (${\rm rank}\left( \mU \right)=1$) as before. The optimization problem~\eqref{eq:update:radar:ris} is  convex and is then solved using off-the-shelf convex solvers. Let $\tilde{\mU}$ be the solution to~\eqref{eq:update:radar:ris}. The required rank-1 solution is then extracted from $\tilde{\mU}$ using Gaussian randomization (refer to Sec.~\ref{design1:step2}).
	
	To ensure that sufficient power is radiated towards different target directions, we initialize radar-\ac{ris} phase shifts so that equally powerful beams are formed towards all target directions. The selection and update of the target \ac{sinr} $\Gamma^{(n)}$ at each iteration is as before. The proposed procedure to solve $(
	\mathcal{P}2)$ is summarized as Algorithm~\ref{Alg_2}.

	\begin{algorithm}[!t] 
		\caption{Solver for $(\mathcal{P}2)$}\label{Alg_2}
			{\bf Initialization:} $\Gamma^{(n)} = \Gamma$, $\boldsymbol{\omega}_{\rm c}^{(n)} = \boldsymbol{\omega}_{\rm c}^{(0)}$, $\boldsymbol{\omega}_{\rm r}^{(n)} = \boldsymbol{\omega}_{\rm r}^{(0)}$
		\begin{algorithmic}[1]
			\For{$n=1, 2, \cdots,\texttt{MaxIter}$}
			\State Solve~\eqref{sp3} with SINR constraint set to $\Gamma^{(n)}$ to update $\hat{\mR}$ and $\{\tilde{\mC}_1,\tilde{\mC}_2, \cdots,\tilde{\mC}_K\}$. 
			\State Update $\{\hat{\mC}_1,\hat{\mC}_2, \cdots,\hat{\mC}_K\}$ using~\eqref{sp3_c_construct} .
			\State Update $\hat{\mS}$ using~\eqref{sp3_s_construct}.
			\State Update $\boldsymbol{\omega}_{
				\rm c}$ as in~\eqref{eq:omega_c_update} and compute $\Gamma_3^{(n)}$. 
			\State Update $\boldsymbol{\omega}_{
				\rm r}$ by solving~\eqref{eq:update:radar:ris}. 
			\If{$\Gamma_3^{(n)} \geq \Gamma $} \texttt{break} \EndIf
			\State Set $\Gamma^{(n)} = \Gamma_2^{(n)} $.
			\EndFor	
		\end{algorithmic}
	\end{algorithm}
	
	\subsubsection{Computational complexity} \label{sec:dual_ris:initialization}
	Similar to the update of comm-RIS phase shifts, updation of radar-RIS phase shifts also involves solving an SDP with an associated complexity of $\cO\left( N^{6.5} \log (1/\epsilon)\right)$.  Hence, the overall complexity is about $\cO \left( I_{\rm out} K^{6.5}M^{6.5}\log\left(1/\epsilon\right)  + I_{\rm out}(I_{\rm in}+1) N^{6.5} \log \left(1/\epsilon \right)\right)$ flops, which is of the same order as that of Algorithm~\ref{Alg_1}.

	\section{Numerical simulations} \label{sec:simulations}
	
			\begin{table}[t]
			\begin{center}
				\begin{tabular}{ |l|l| } 
					\hline
					Parameter & Value \\
					\hline  
					\ac{dfbs} location & $\left(0,0,0\right)$ m	\\	
					comm-\ac{ris} location & $\left(20,13,3\right)$ m	\\	
					radar-\ac{ris} location & $\left(-6,6,3\right)$ m	\\	
					Target angles w.r.t. \ac{dfbs} & $\left(-70^\circ,-50^\circ,-30^\circ,-20^\circ,-10^\circ\right)$\\
					Target distance from \ac{dfbs} & $5$ m \\

					\hline
				\end{tabular}
			\end{center}
			\caption{Simulation parameters.}
			\label{table:simparam_locations}
		\end{table}
		
		\begin{figure*}[ht!]
		\begin{subfigure}[c]{0.48\columnwidth}\centering
			\includegraphics[width=1\columnwidth]{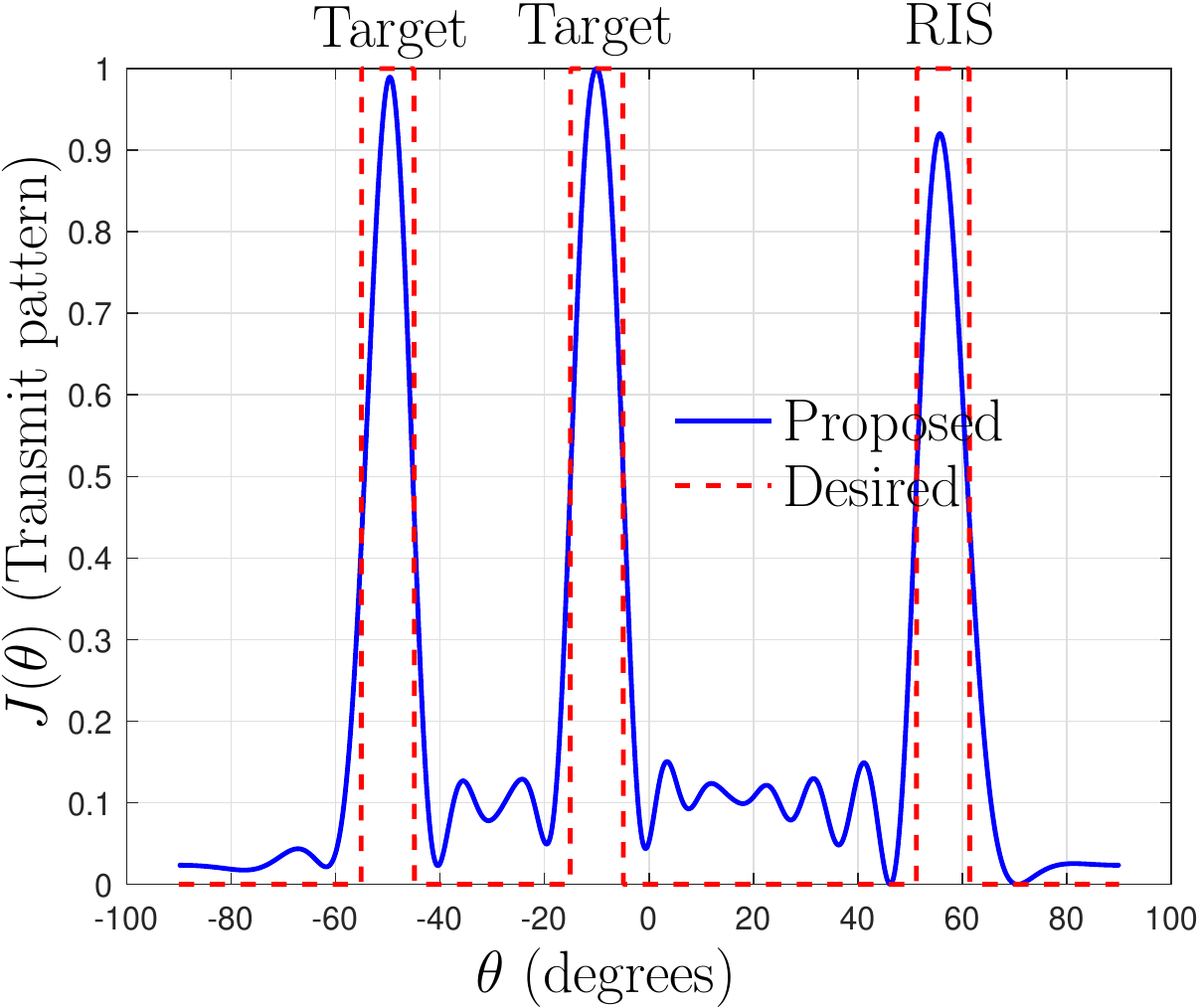}
			\caption{}
			\label{fig:a1_tx}
		\end{subfigure}
		~
		\begin{subfigure}[c]{0.48\columnwidth}\centering
			\includegraphics[width=1\columnwidth]{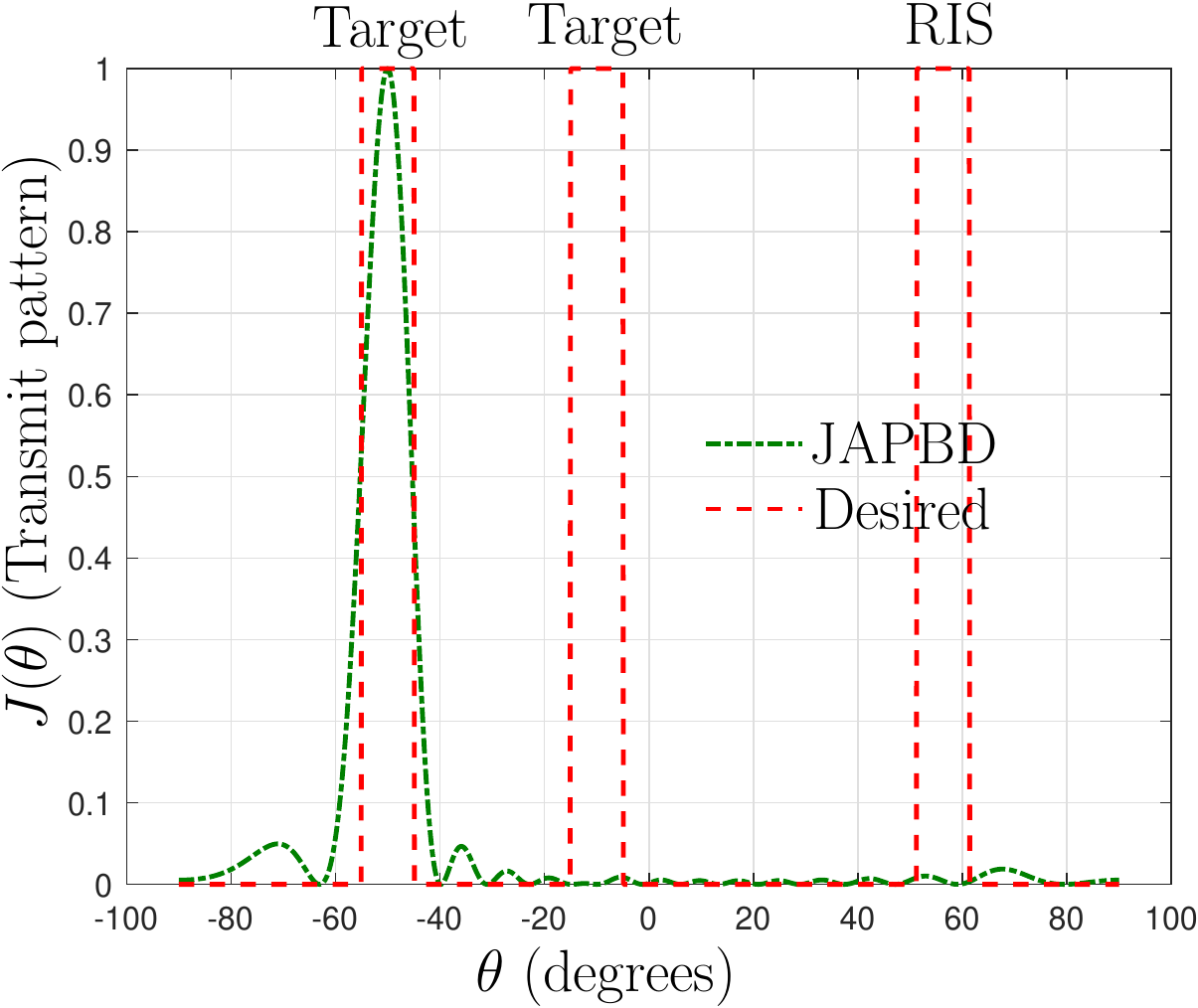}
			\caption{}
			\label{fig:a1_tx}
		\end{subfigure}
		~
		\begin{subfigure}[c]{0.48\columnwidth}\centering
			\includegraphics[width=1\columnwidth]{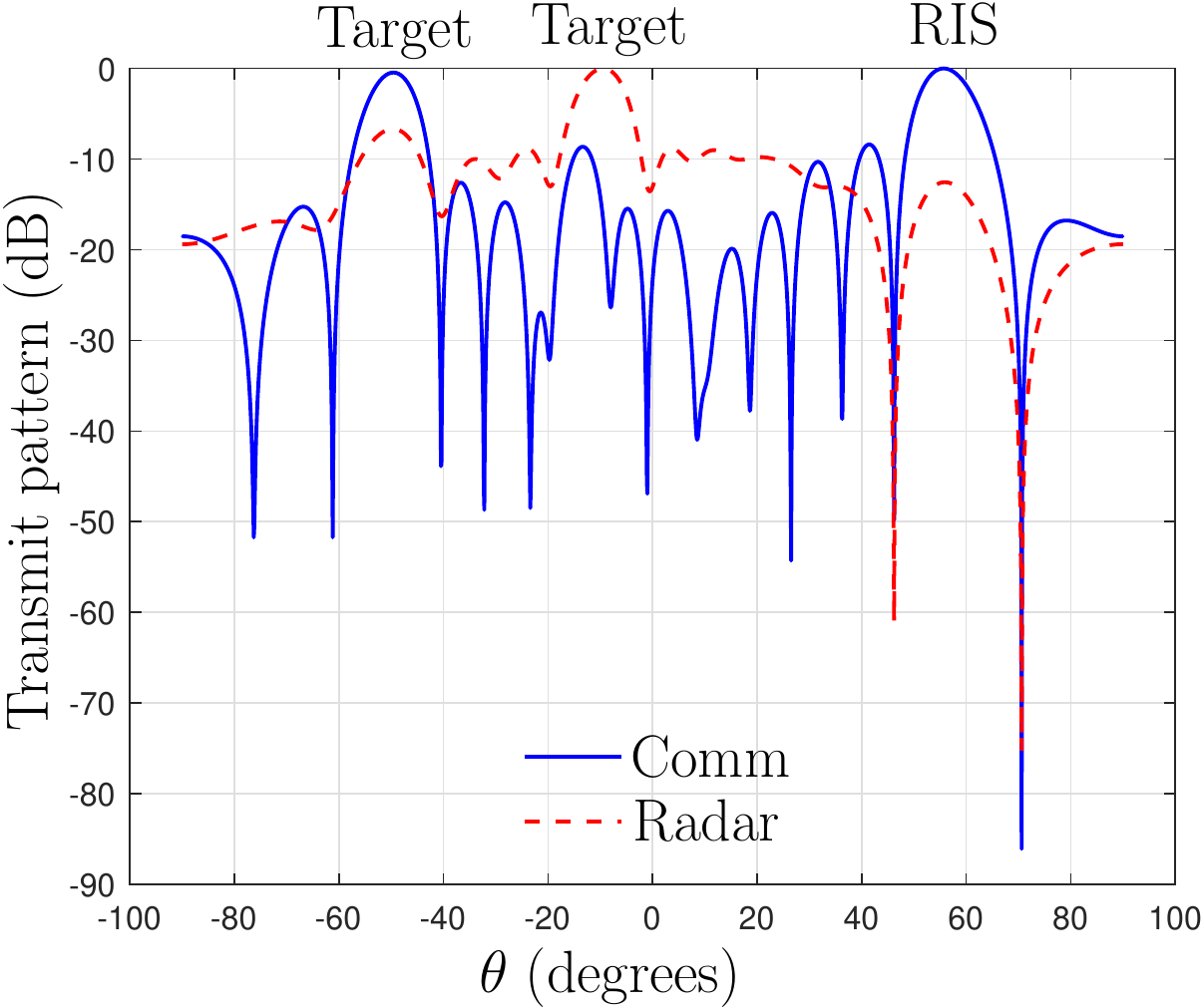}
			\caption{}
			\label{fig:a1_tx_split}
		\end{subfigure}
		~
		\begin{subfigure}[c]{0.48\columnwidth}\centering
			\includegraphics[scale=0.4]{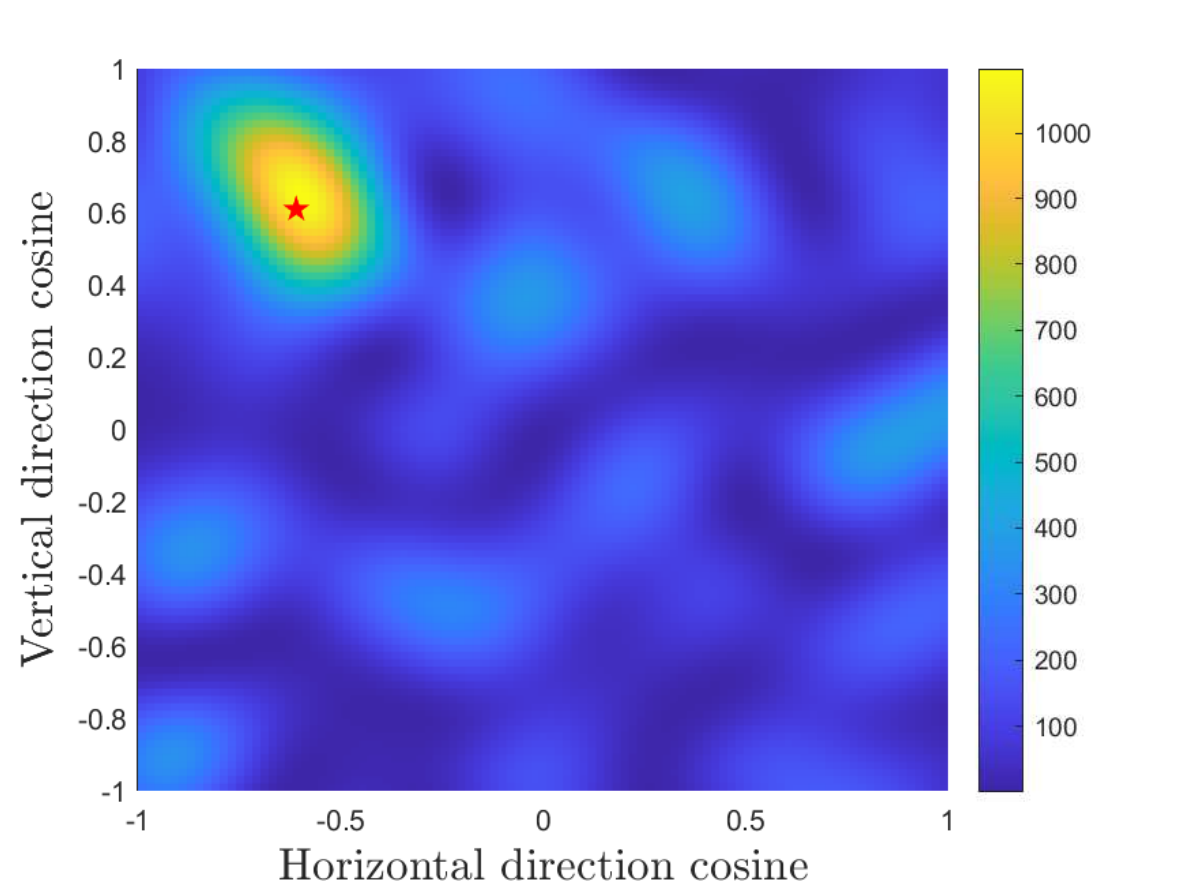}
			\caption{}
			\label{fig:a1_ris}
		\end{subfigure}
		\caption{ (a) Transmit beampattern at the \ac{dfbs}  of the proposed method (b) Transmit beampattern at the \ac{dfbs} of JAPBD~\cite{hua2022joint}   (c) Sensing and communication transmit beams viewed separately. (d) Reflection profile at the comm-\ac{ris}. Star~($\star$) indicates the true location of the \ac{ue} w.r.t. the \ac{ris}.}
		\label{fig:alg1_beams}
	\end{figure*}

	In this section, we present results from a number of numerical experiments to illustrate the benefits of \ac{ris}-enabled \acs{isac} systems and performance of the proposed algorithms. Throughout the simulations, we model the \ac{dfbs} as a \ac{ula} with $M=16$ half-wavelength spaced elements. Both the comm- and radar-\acp{ris} are modeled as \acp{ura} comprising of quarter-wavelength spaced elements~\cite{renzo2020smart}. We model the entries of the channels $\mH_{\rm br}$, $\mG_{\rm br}$, and $\vh_{{\rm ru},k}\rH$ for $k=1,\ldots,K$ as Rician distributed random variables  with a Rician factor of $\rho = 10$. The direct channel for the communication \acp{ue}, i.e., $\vh_{{\rm bu},k}\rH$ for $k=1,\ldots,K$ is assumed to undergo Rayleigh fading. The remaining channels, namely, 
	$\vg_{{\rm bt},m}$ and $\vg_{
		{\rm rt},m}$ for $m=1,\ldots,T$ are modeled as \ac{los} channels.

	The receiver noise variance at the \acp{ue} is set to $-94$ dBm.  The targets are assumed to be located in the far-field of both the \ac{dfbs} and radar-\ac{ris}.  User locations are drawn randomly from a rectangular grid having corners $(15,8,0)$m, $(15,18,0)$m, $(18,8,0)$m, and $(18,18,0)$m. The pathlosses of  radar-\ac{ris} and target links and \ac{dfbs} and \ac{ue} links are modeled as $30 + 25 \log d$	dB and $30 + 36 \log d$	dB , respectively. Pathlosses for the rest of the links are modeled as $30 + 22 \log d$	dB with $d$ being the distance between  concerned terminals in m.  Rest of the simulation parameters are presented in Table~\ref{table:simparam_locations}.

	\subsection{Comm-RIS assisted ISAC system}  \label{sec:simulations:comm_ris}
	We begin by presenting the designed transmit beampattern and the reflection profile of the comm-\ac{ris}. Then we present results from the Monte-Carlo experiments to show the benefits of \ac{ris} in \acs{isac} systems in terms of the achievable \ac{sinr} and worst-case  illumination power at the target locations of interest by varying different system parameters such as \ac{sinr} requirement ($\Gamma$),  number of \ac{ris} elements ($N$), and number of users with direct path ($K_d$). In this section, we consider a two-target scenario, i.e., $T=2$ with $\theta_1 = -50^\circ$ and $\theta_2 = -10^\circ$. 
	
	We compare the performance of the proposed method in terms of transmit pattern, the fairness \ac{sinr} (or min-rate), and the worst-case target illumination power with the following four schemes: (\emph{i}) joint active and passive beamforming design~(\texttt{JAPBD}) in~\cite{hua2022joint}, (\emph{ii}) \texttt{\ac{ris} manual selection}, wherein we select the \ac{ris} phase profile to form equally powerful beams towards all the \acp{ue}, and (\emph{iii}) \texttt{no RIS}, wherein we consider an \acs{isac} system without \ac{ris}~\cite{liu2020joint_transmit_beamform}. We compute the beamformers at the \ac{dfbs} by solving~\eqref{sp3} for both \texttt{\ac{ris} manual selection} and \texttt{no RIS} schemes. Although such manual selection of \ac{ris} phase shifts seem intuitive, they are agnostic to the task at hand and are not fairness \ac{sinr} optimal. We also compare the proposed method with (\emph{iv}) \texttt{sensing-only system}, which does not have any communication functionality or \ac{ris}~\cite{stoica2007on_probing_signal}. For the \texttt{sensing-only system}, we solve $(\mathcal{P}1)$ only w.r.t. ~$\mS$ ($\mC$ is set to zero) without the  constraints \eqref{problem_formulation_sinr} and \eqref{problem_ris_constraint}. To simulate different schemes, we first run \texttt{JAPBD} to compute the minimum total transmit power required to achieve certain radar \ac{snr}, communication \ac{sinr}, and  average squared cross-correlation between the beams. We then choose the resulting power as the total power constraint for the remaining schemes, thereby ensuring that all schemes utilize the same amount of transmit power.

	\subsubsection{Transmit beampattern and \ac{ris} reflection pattern}

		\begin{figure}[t]
		\centering
		\includegraphics[width=0.8\columnwidth]{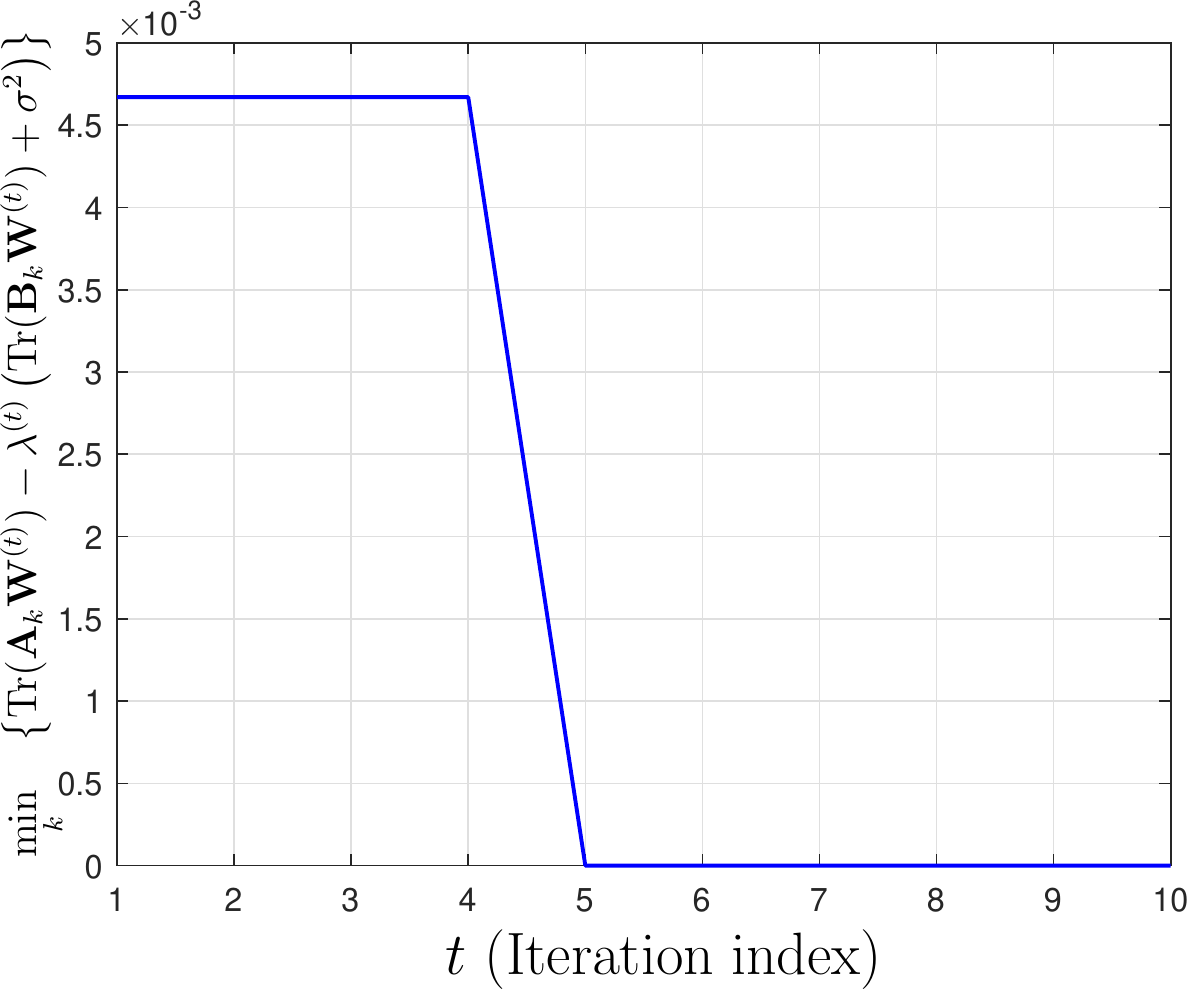}
		
		\caption{ Convergence of Dinkelbach-like iterations for updating~$\boldsymbol{\omega}_{\rm c}$.}
		\label{fig:a1_p2_converge}
	\end{figure}

	For the ideal beampattern $d(\theta)$ in~\eqref{d_theta_design}, we use a superposition of rectangular box functions of width $\epsilon = 10$ degrees. We set  $w_b = w_c = 1$ in $L(\mR,\tau)$~[cf. \eqref{eq:radar_cost1}]. To illustrate the beampattern obtained from Algorithm~\ref{Alg_1}, we consider $K=1$. We use $N = 100$ \ac{ris} elements and $\Gamma = 5~{\rm dB}$. The transmit beampattern at the \ac{dfbs} is shown in Fig.~\ref{fig:alg1_beams}(a) and \ref{fig:alg1_beams}(b). The transmit pattern of a scheme that considers sum received \ac{snr} as the radar metric~\cite{hua2022joint} results in a peak only towards one of the targets. As a consequence, a system designed using the metric presented in~\cite{hua2022joint} completely misses one of the targets. On the other hand, the transmit beampattern of the proposed algorithm  has a response towards the two target locations of interest as well as towards the comm-\ac{ris}, as desired, thereby ensuring that none of the targets are missed.

	\begin{figure*}[ht]
		\begin{subfigure}[c]{0.64\columnwidth}\centering
			\includegraphics[width=\columnwidth]{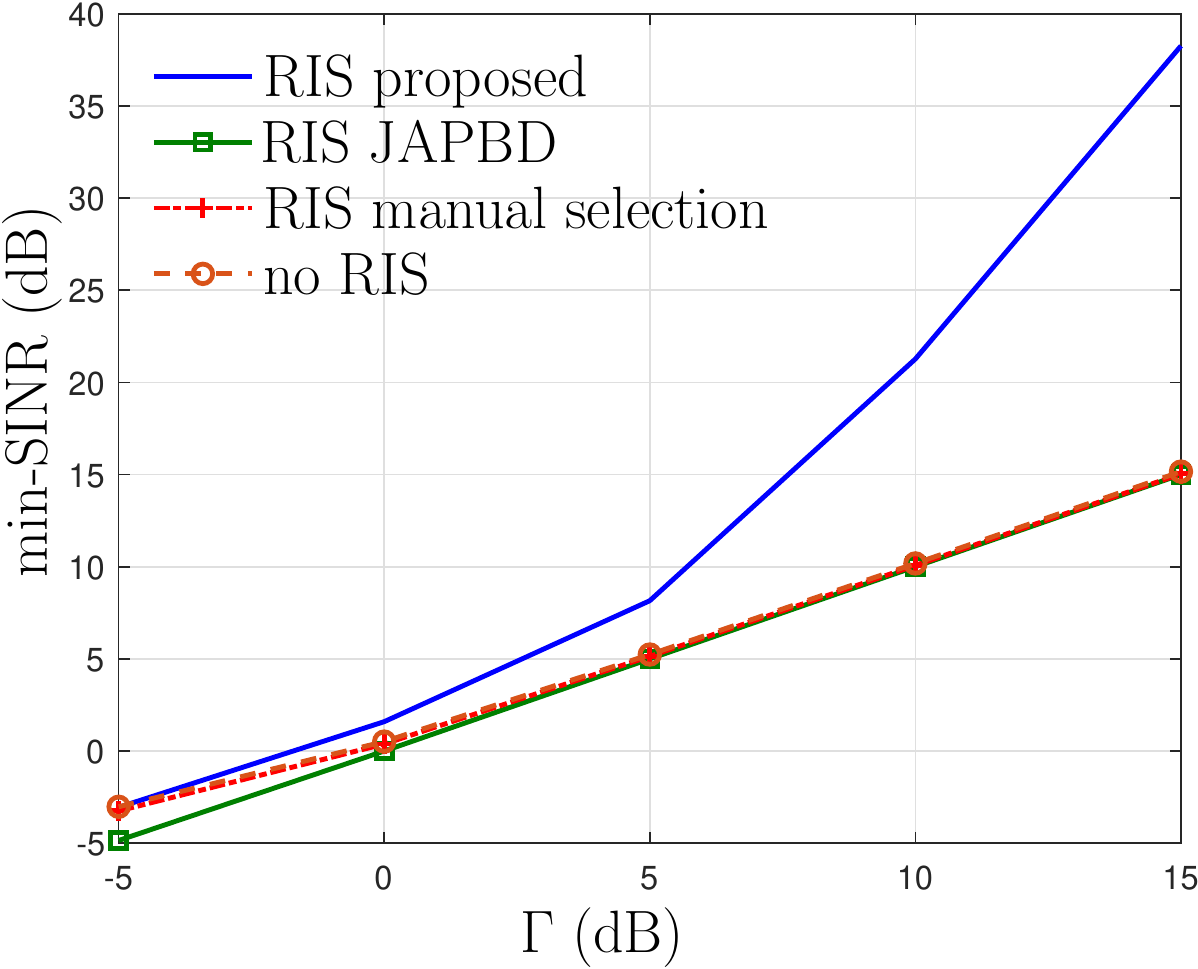}
			\vspace{2mm}
			\caption{}
			\label{fig:a1_sinr_gamma}
		\end{subfigure}
		~
		\begin{subfigure}[c]{0.64\columnwidth}\centering
			\includegraphics[width=\columnwidth]{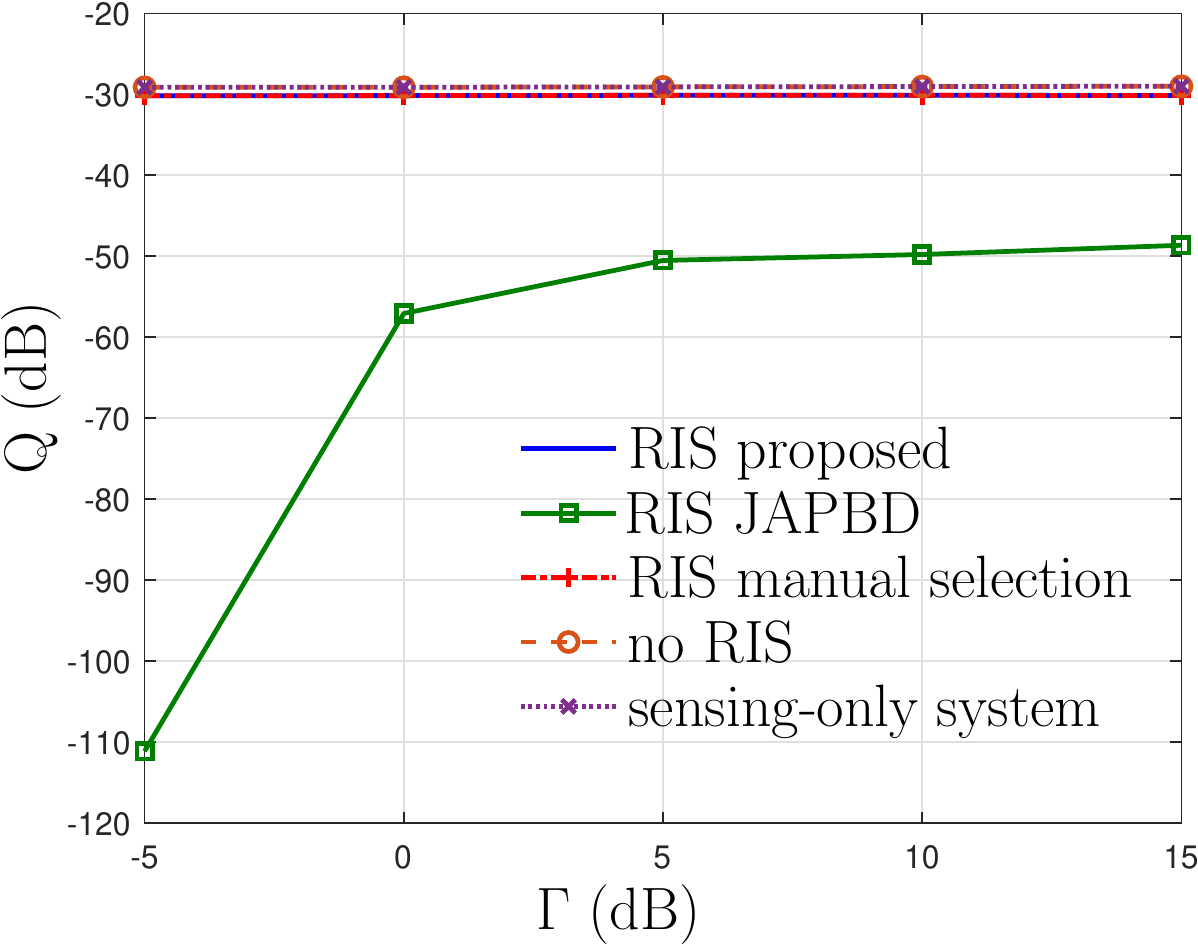}
				\vspace{2mm}
			\caption{}
			\label{fig:a1_q_gamma}
		\end{subfigure}
		~
		\begin{subfigure}[c]{0.64\columnwidth}\centering
			\includegraphics[width=\columnwidth]{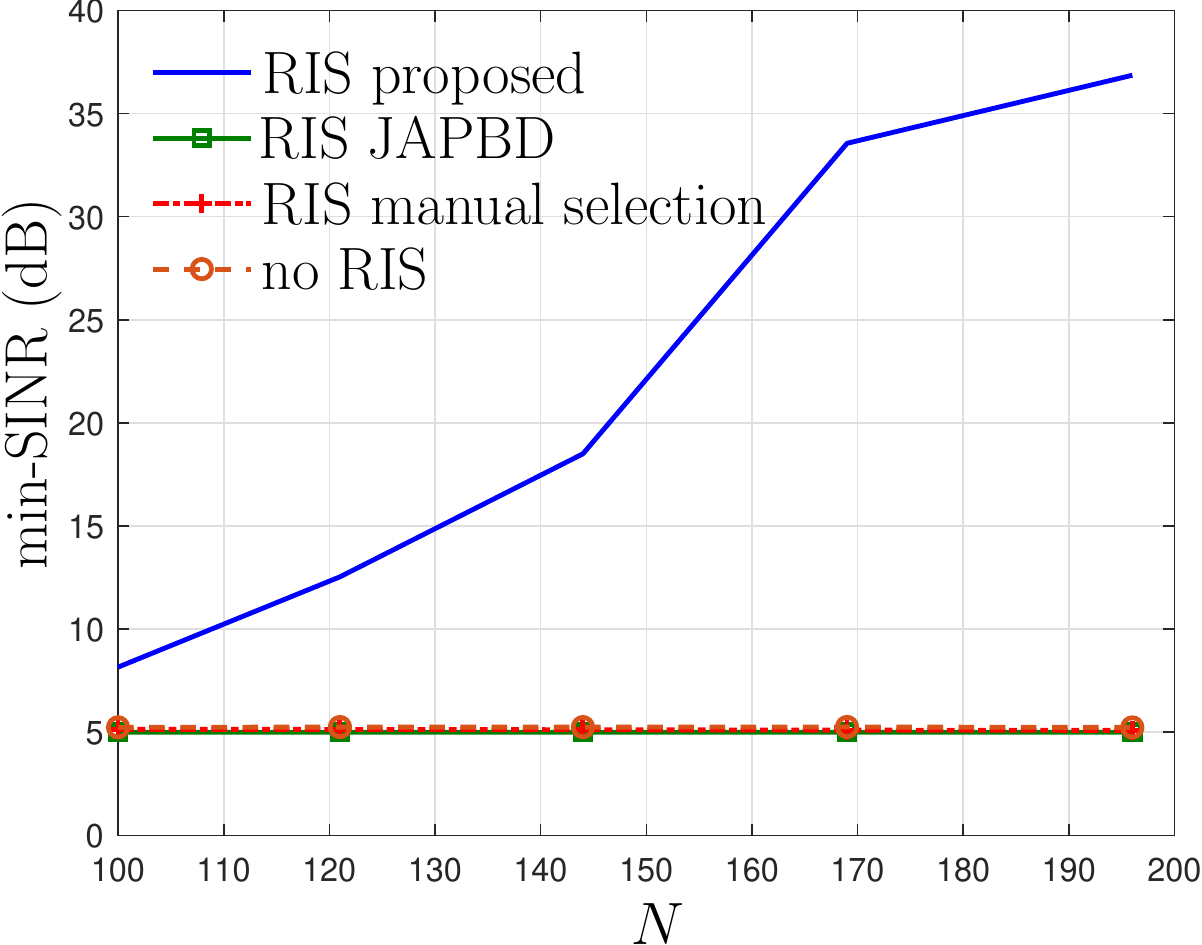}
				\vspace{2mm}
			\caption{}
			\label{fig:a1_sinr_N}
		\end{subfigure}
		~ 
		\begin{subfigure}[c]{0.64\columnwidth}\centering
			\includegraphics[width=\columnwidth]{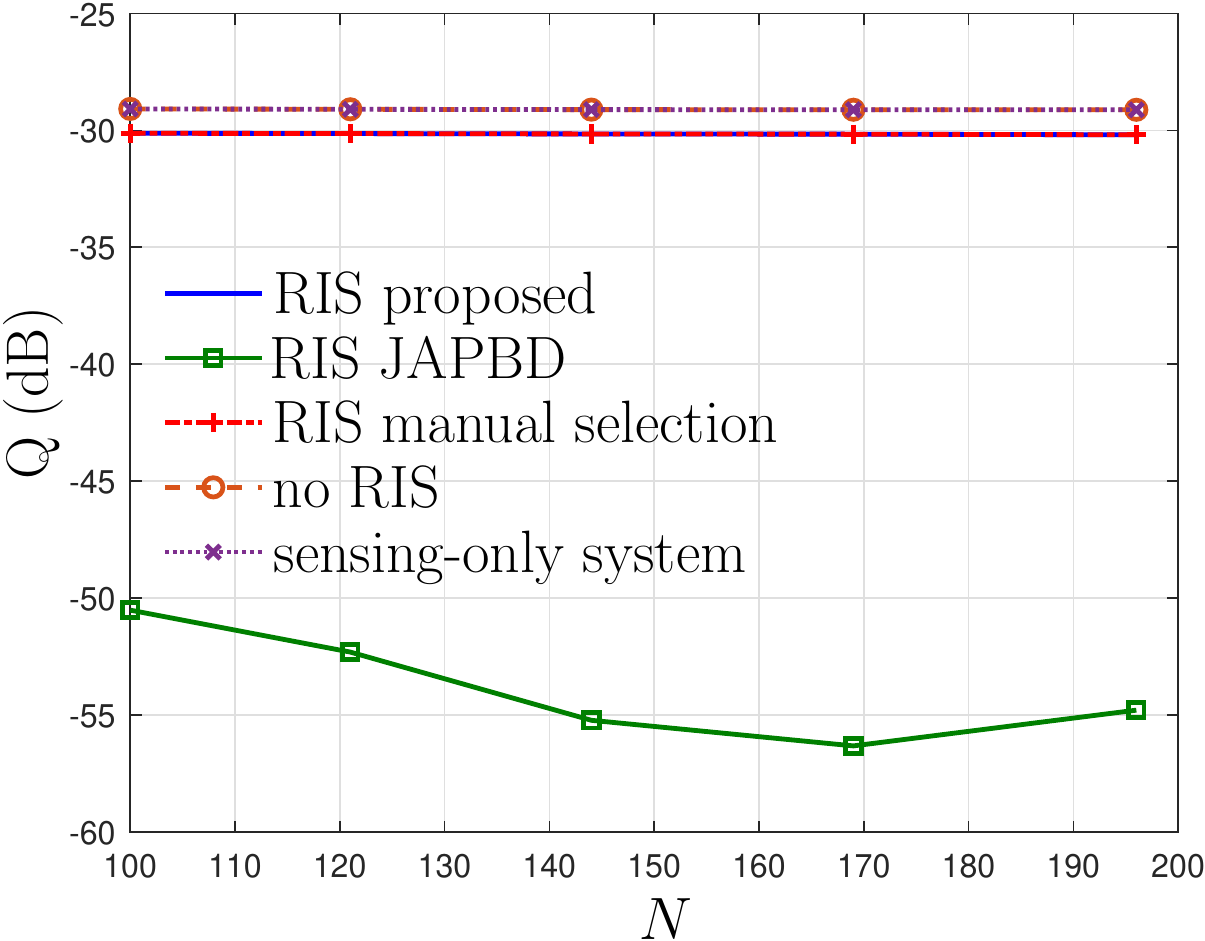}
			\caption{}
			\label{fig:a1_q_N}
		\end{subfigure}
		~~~
		\begin{subfigure}[c]{0.64\columnwidth}\centering
			\includegraphics[width=\columnwidth]{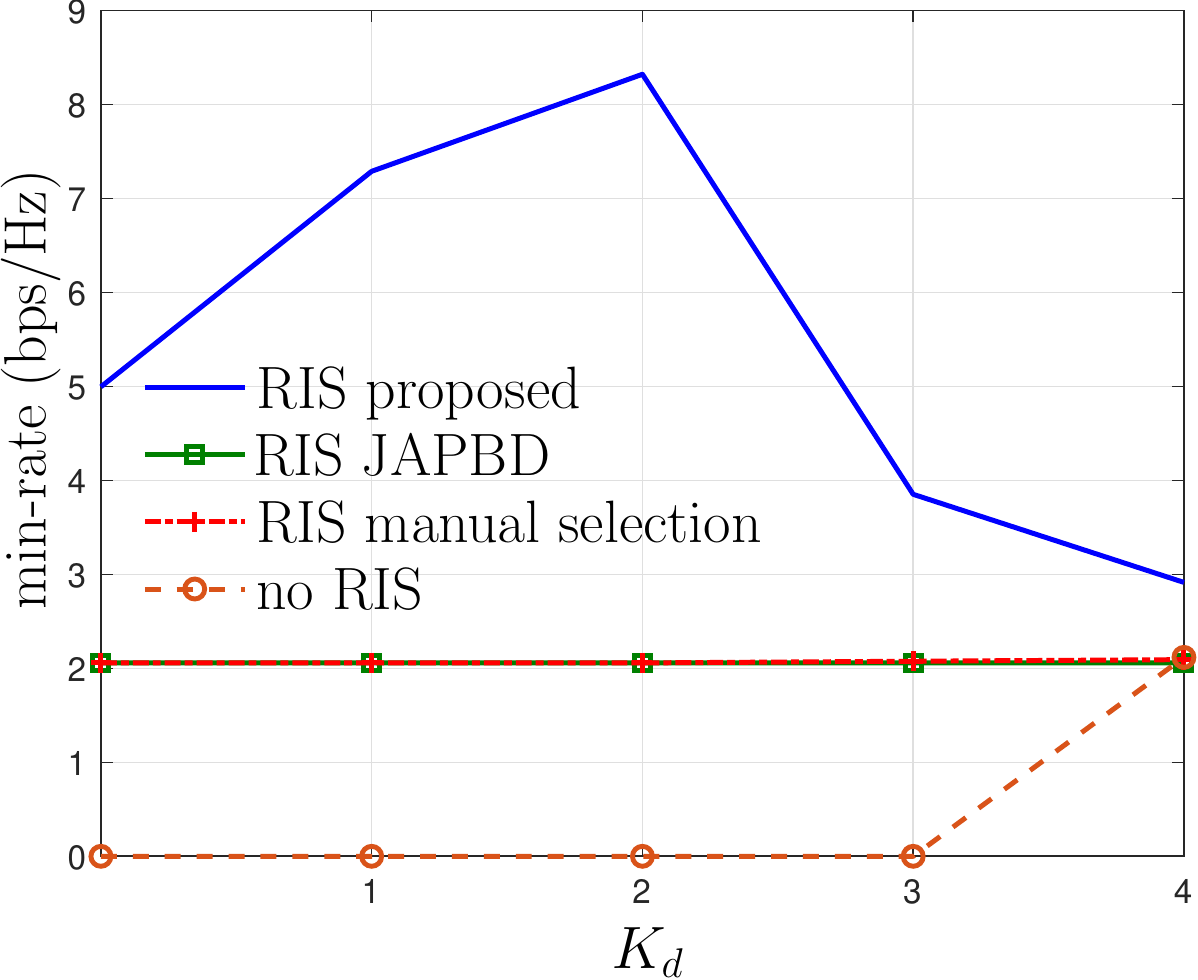}
			\caption{}
			\label{fig:a1_minrate_Nd}
		\end{subfigure}
		~
		\begin{subfigure}[c]{0.64\columnwidth}\centering
			\includegraphics[width=\columnwidth]{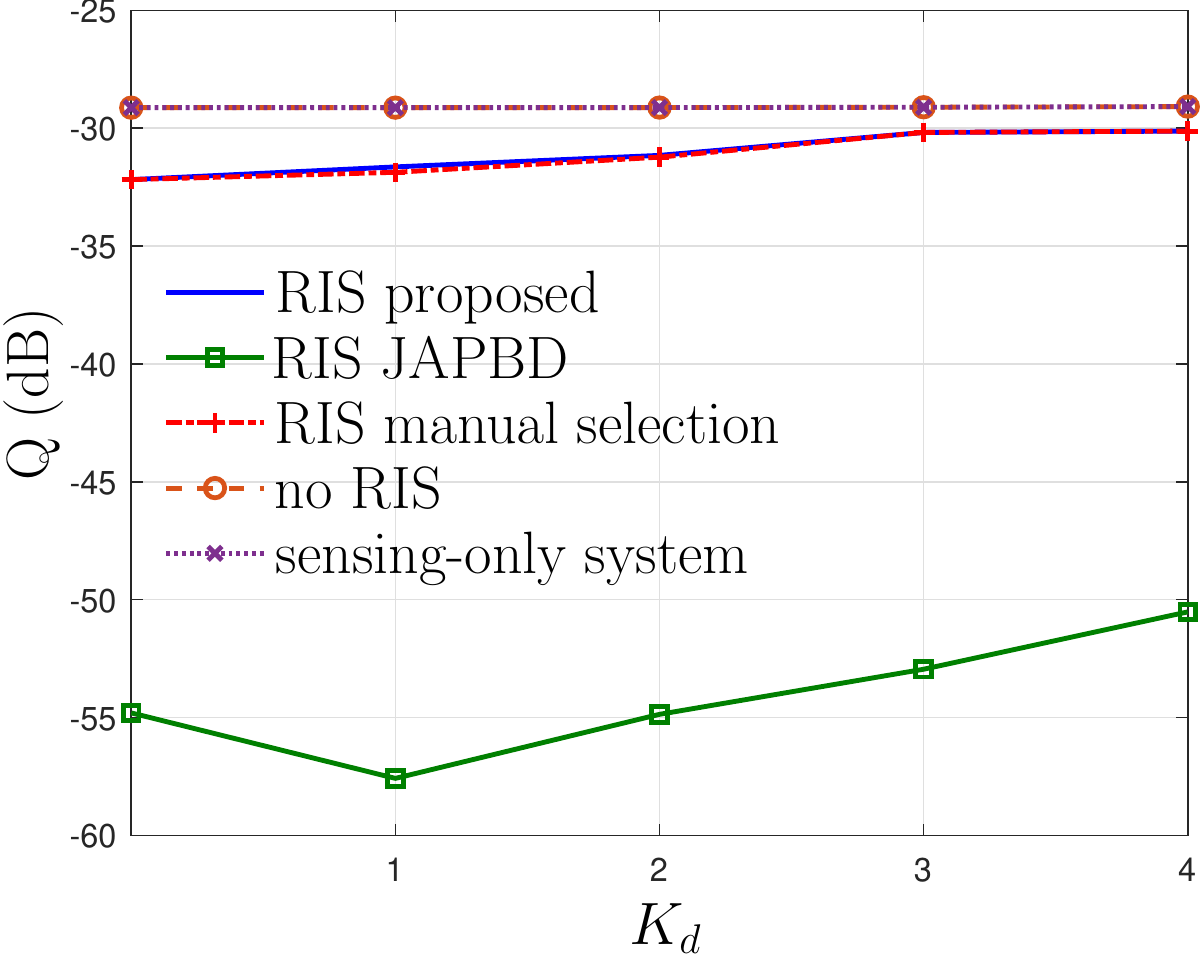}
			\caption{}
			\label{fig:a1_q_Nd}
		\end{subfigure}
		\caption{ Comm-\ac{ris} assisted \acs{isac} system. Impact of $\Gamma$ on (a) fairness \ac{sinr} and (b) worst-case target illumination power. Impact of $N$ on (c) fairness \ac{sinr} and (d) worst-case target illumination power. Impact of $K_d$ on (e) min-rate and (f) worst-case target illumination power.  }
		\label{fig:alg1_sim}
	\end{figure*}

	To gain further insights, we consider individual beampatterns of the radar and communication beamformers, where the radar beampattern is defined as
	$
	J_{R} (\theta) = \va\rH(\theta) \mS\mS\rH\va(\theta)
	$
	and the beampattern related to the $k$th communication user is defined as
	$
	J_{C,k}(\theta)  = \va\rH(\theta) \vc_k \vc_k\rH \va(\theta).
	$
	To visualize the reflection pattern at RIS, we define the beampattern of RIS towards the direction cosine vector ${\boldsymbol \psi} \in \mathbb{R}^2$ for a signal incident on the RIS from the \ac{dfbs} from the direction ${\boldsymbol \phi} \in \mathbb{R}^2$ as follows
	$
	J_{\rm RIS}({\boldsymbol \psi}) = \vert  \vr\rH({\boldsymbol \psi}) {\rm diag}(\boldsymbol{\omega}) \vr({\boldsymbol \phi})    \vert,
	$
	where $\vr(\cdot) \in \mbC^{N \times 1}$ is the array response vector of the \ac{ris}.
	The separate beampatterns (related to the sensing and communication beamformers) at the \ac{ris} is shown in Fig.~\ref{fig:alg1_beams}(c). We observe that the radar beam has peaks towards target directions of interest and the communication beam has a stronger peak towards the \ac{ris}. Further, the radar beam has a dip in the direction of the \ac{ris}. This is desired since the comm-\ac{ris} is used solely to serve the \acp{ue}, and any amount of radar signals transmitted towards the \acp{ue} via the \ac{ris} could lead to potentially increased interference~(hence lower SINR) at the communication \acp{ue}. While transmitting radar signals to communication \acp{ue} leads to interference, transmitting communication signals (known at the \ac{dfbs}) to target locations actually helps in improving the target illumination power, and this explains the peaks of communication beams towards the targets. In Fig.~\ref{fig:alg1_beams}(d), we show the reflection pattern of the \ac{ris}, where we can see that the \ac{ris} beampattern has a peak towards the user direction. This is intuitive since the \ac{ris} attempts to steer all the energy incident on it towards the single \ac{ue} to maximize the received \ac{sinr}. For larger values of $K$, unlike the $K=1$ case, the reflection pattern at the \ac{ris} is often not interpretable as the \ac{sinr} optimal design is not necessarily the pattern that forms beams towards $K >1$ users.
	
	In Fig.~\ref{fig:a1_p2_converge}, we show convergence of the updates~\eqref{eq:dinkelbach_1}-\eqref{eq:dinkelbach_2} that solve \eqref{subproblem2_mod3} for updating the comm-\ac{ris} phase shifts. Here, we use a particular channel realization with $K=4$ (This means we have a fractional programming problem with four ratios).

	\subsubsection{Monte-Carlo simulations}
	
	We now consider a multi-user, multi-target scenario with $K=4$, $T=2$, and $N=100$, unless mentioned otherwise.  We present a number of results from Monte-Carlo experiments that are obtained by averaging over 100 independent realizations of the fading channel with different user locations.  Unless mentioned otherwise, we use radar \ac{snr} requirement of $10$ dB and a average squared cross-correlation of $10^{-5}$.

	In Fig.~\ref{fig:alg1_sim}(a), we show the impact of \ac{sinr} constraint ($\Gamma$) on the fairness \ac{sinr} of different methods. While the benchmark schemes provide exactly the minimum required \ac{sinr}, the proposed method provides significantly high \ac{sinr}, without consuming any additional power. This is due to the fact that the selection of comm-\ac{ris} phase shifts is carried out to maximize the \ac{sinr} itself, unlike the other schemes.  Impact of $\Gamma$ on the worst-case target illumination power is presented in Fig.~\ref{fig:alg1_sim}(b). The worst-case target illumination power of \texttt{JAPBD} is significantly less than that of other schemes. This happens due to the choice of sum radar \ac{snr} as the performance metric in \texttt{JAPBD} which results in scenarios like the one presented in Fig.~\ref{fig:alg1_beams}(a) wherein not all targets are  illuminated. On the other hand, due to the use of a weighted sum of beampattern mismatch error and average squared cross-correlation as the radar metric, the radar performance of the rest of the schemes are much higher than that of \texttt{JAPBD}. Moreover, performance of the proposed scheme is also comparable to that of a \texttt{sensing-only system}, which is the benchmark method since an \ac{isac} system cannot achieve better sensing performance than a comparable sensing-only system. The worst-case target illumination power of the proposed method is only about $1.5$ dB worse than that of an \ac{isac} system without the \ac{ris}. This degradation is due to the fact that we form a beam towards the \ac{ris} in addition to the beams towards the target directions of interest. We can also observe that the worst-case target illumination power of  \texttt{no RIS} and \texttt{sensing-only system} are comparable as these do not form any beam towards the \ac{ris}.

	In Fig.~\ref{fig:alg1_sim}(c) and Fig.~\ref{fig:alg1_sim}(d), we present the  communication and radar performance of different methods for varying number of \ac{ris} elements. As before, \ac{sinr} of the proposed method is significantly higher than that of other benchmark schemes. The worst-case target illumination power  of the proposed scheme is also remarkably better than that of \texttt{JAPBD} and is comparable to that of the benchmark scheme \texttt{sensing-only system}. The fairness \ac{sinr} of the proposed method also increases with an increase in the number of comm-\ac{ris} elements due to the increased array gain offered by the \ac{ris}. On the other hand, since the comm-\ac{ris} is not involved in sensing, changes in $N$ will not affect the worst-case target illumination power. Albeit a moderate loss of about $1.5$ dB in $Q$ due to the formation of an additional beam towards the \ac{ris}, the fairness \ac{sinr} is significantly improved, often by more than $10$ dB by properly designing the phase profile of the comm-\ac{ris}.

	Next, we consider a setting where only $K_d$ out of $K$ users have a direct path from the \ac{dfbs} and the direct paths to the remaining $K - K_d$ \acp{ue} are blocked. For the considered scenario, an \ac{isac} system without \ac{ris} would only be able to serve $K_d$ users out of $K$. We show the minimum-rate, i.e., $\min_k  \log\left(1 + \gamma_k\right)$ and the worst-case target illumination power for different number of \acp{ue} with direct path, i.e., $K_d$, in Fig.~\ref{fig:alg1_sim}(e) and Fig.~\ref{fig:alg1_sim}(f), respectively.  We can see that with the comm-\ac{ris}, all the $K$ users, irrespective of the presence of a direct path or not, is guaranteed a minimum \ac{sinr} of $\Gamma = 5~{\rm dB}$ (or min-rate of  2.05 bps/Hz). However, without the \ac{ris}, only $K_{\rm d}$ users are served with an SINR higher than $5~{\rm dB}$. Although \texttt{RIS manual selection} and \texttt{JAPBD} also offer a min-rate of $2.05$ bps/Hz, the proposed method offers a significantly higher min-rate of about $4-8$ bps/Hz. As before, the value of $Q$ of \acs{isac} systems with \ac{ris} degrades by about $1.5-2.5$ dB compared to an \acs{isac} system without \ac{ris}. We wish to re-emphasize that \texttt{JAPBD} often results in scenarios where not all targets are properly illuminated, leading to low values of the worst-case target illumination power, as indicated in Fig.~\ref{fig:alg1_sim}(f). On the other hand, the proposed scheme ensures that all targets are illuminated with sufficient power even when the direct path to a few of the users is blocked. 
	Throughout the simulations, we have also observed (not shown here) that the mean squared cross-correlation of the proposed method is comparable, often better, than that of \texttt{JAPBD} scheme.

\begin{figure*}[t]
	\begin{subfigure}[c]{0.64\columnwidth}\centering
		\includegraphics[width=\columnwidth]{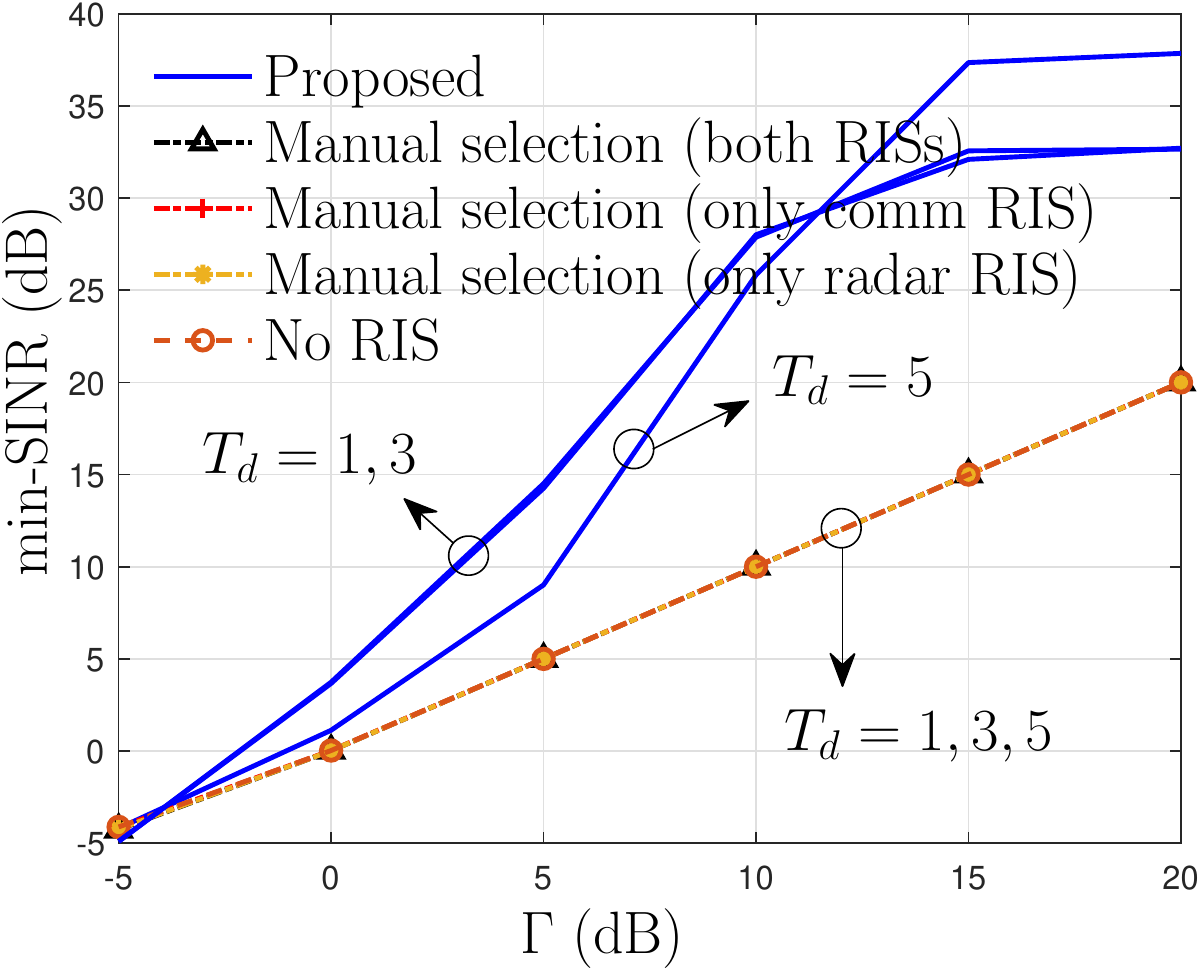}
			\vspace{2mm}
		\caption{}
		\label{fig:a2_sinr_gamma}
	\end{subfigure}
	~
	\begin{subfigure}[c]{0.64\columnwidth}\centering
		\includegraphics[width=\columnwidth]{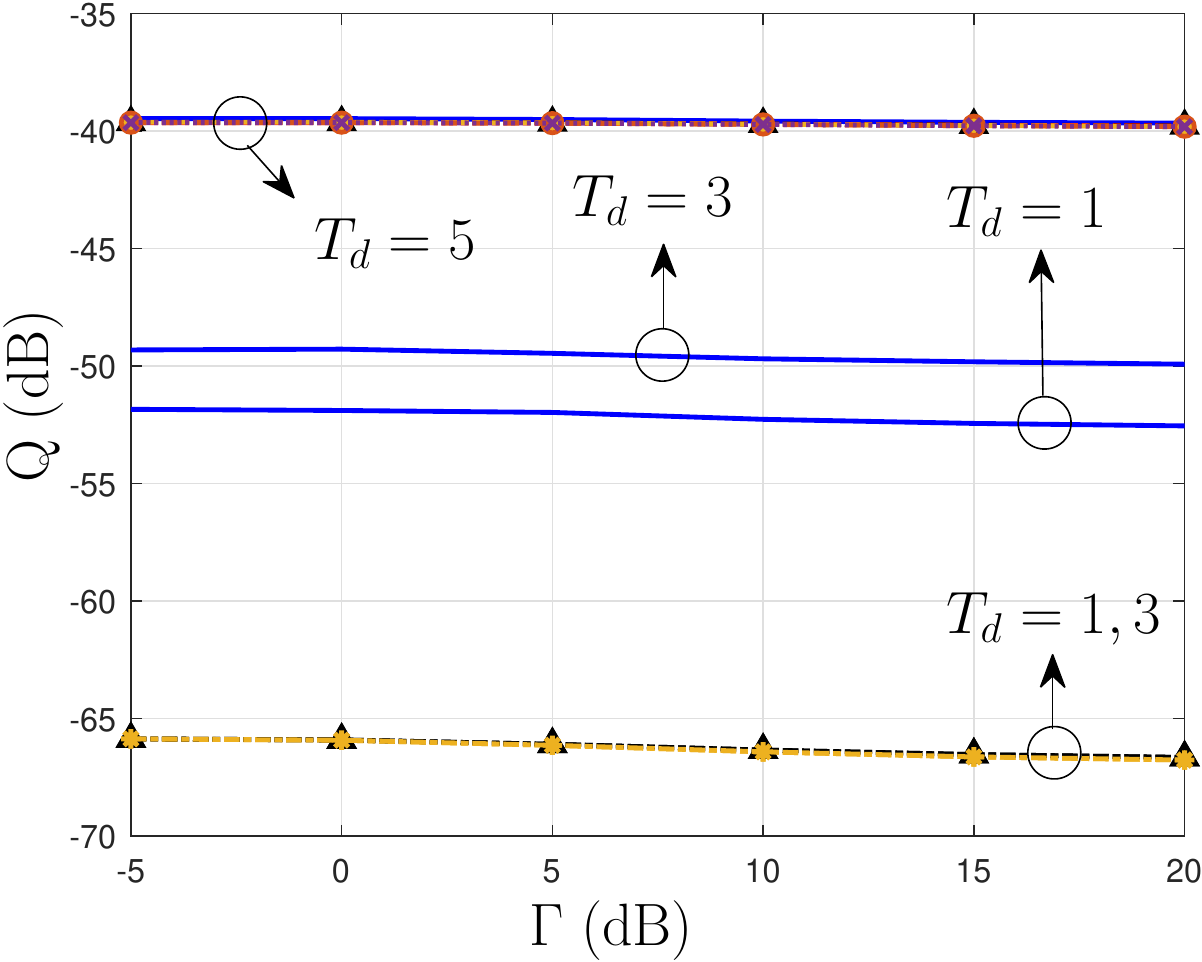}
			\vspace{2mm}
		\caption{}
		\label{fig:a2_q_gamma}
	\end{subfigure}
	~
	\begin{subfigure}[c]{0.64\columnwidth}\centering
		\includegraphics[width=\columnwidth]{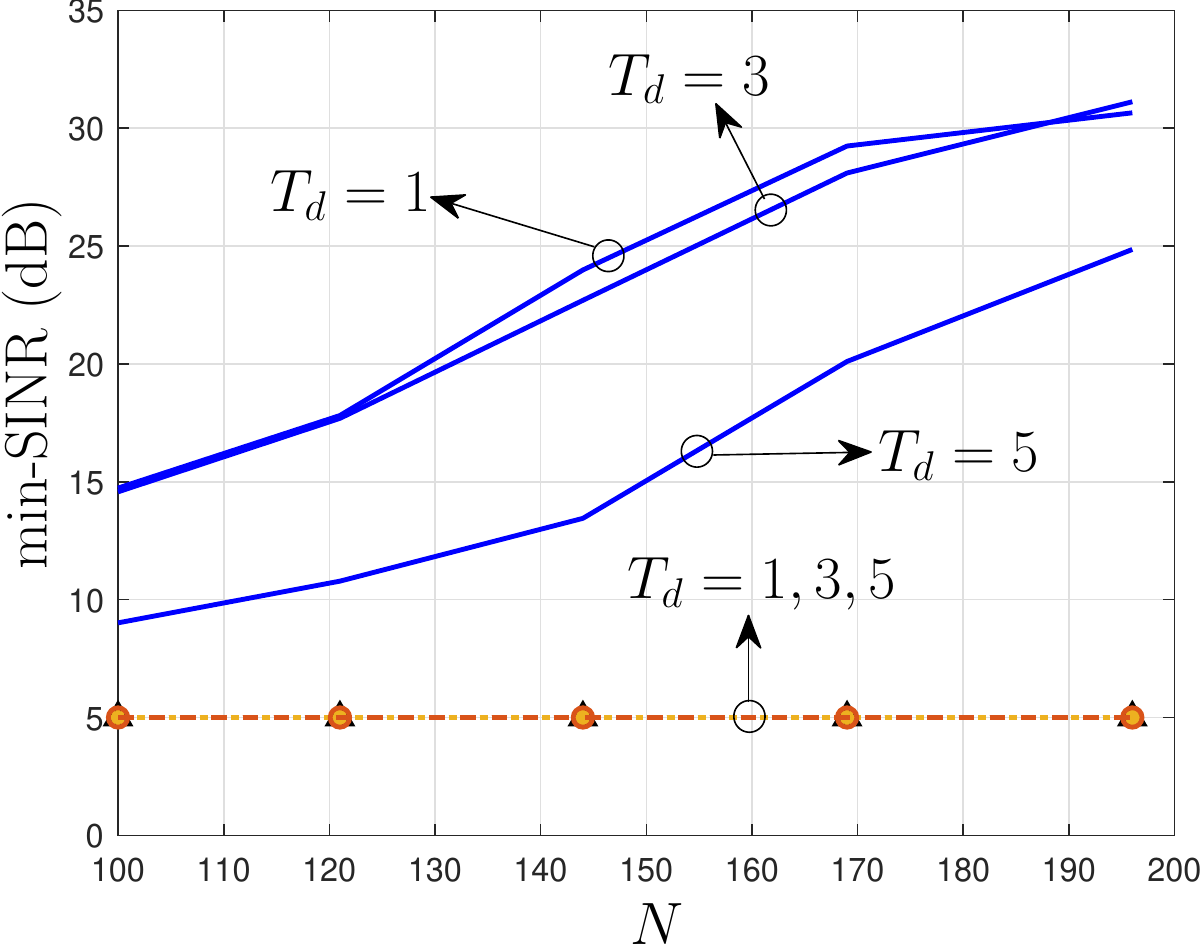}
			\vspace{2mm}
		\caption{}
		\label{fig:a2_sinr_N}
	\end{subfigure}
	~
	\begin{subfigure}[c]{0.64\columnwidth}\centering
		\includegraphics[width=\columnwidth]{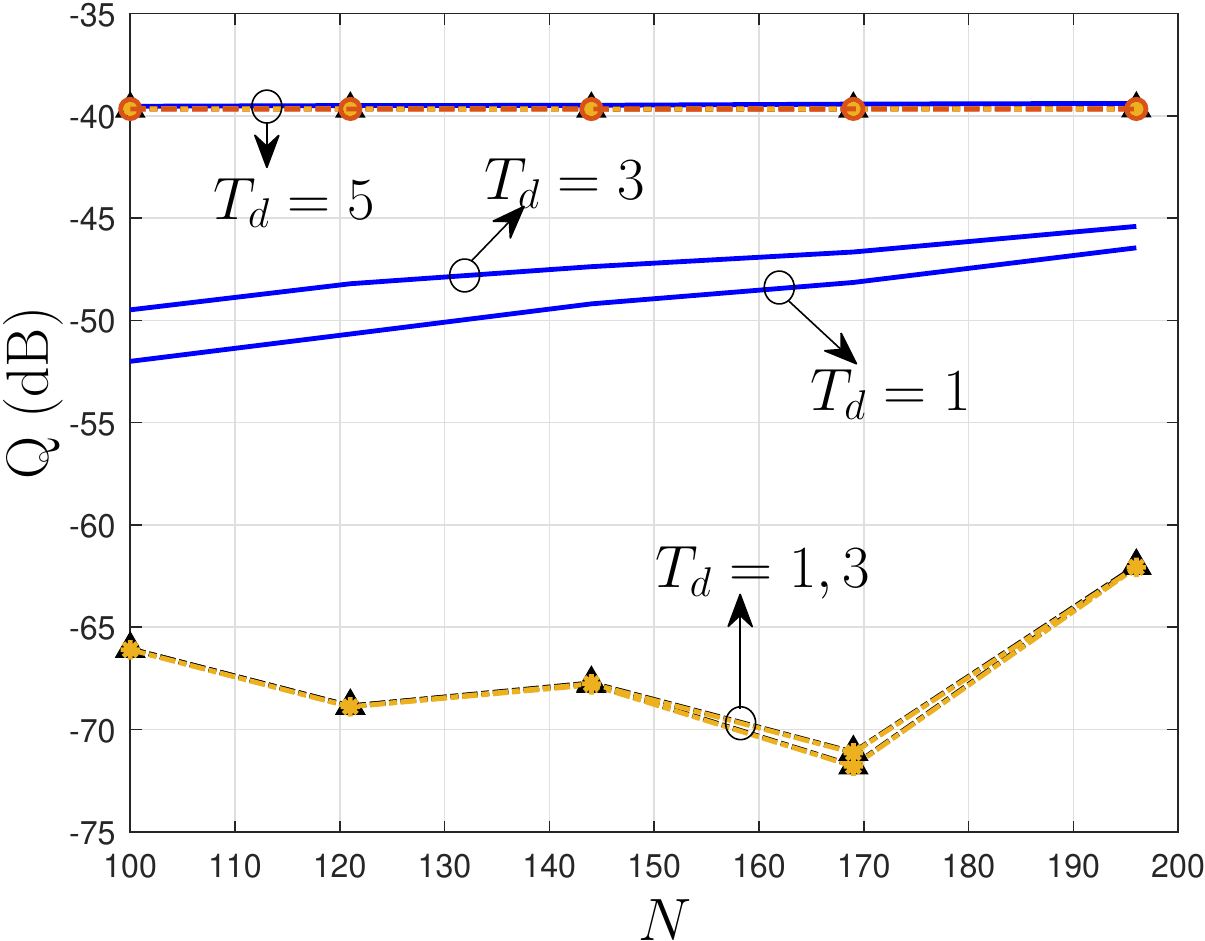}
		\caption{}
		\label{fig:a2_q_N}
	\end{subfigure}
	~~~
	\begin{subfigure}[c]{0.64\columnwidth}\centering
		\includegraphics[width=\columnwidth]{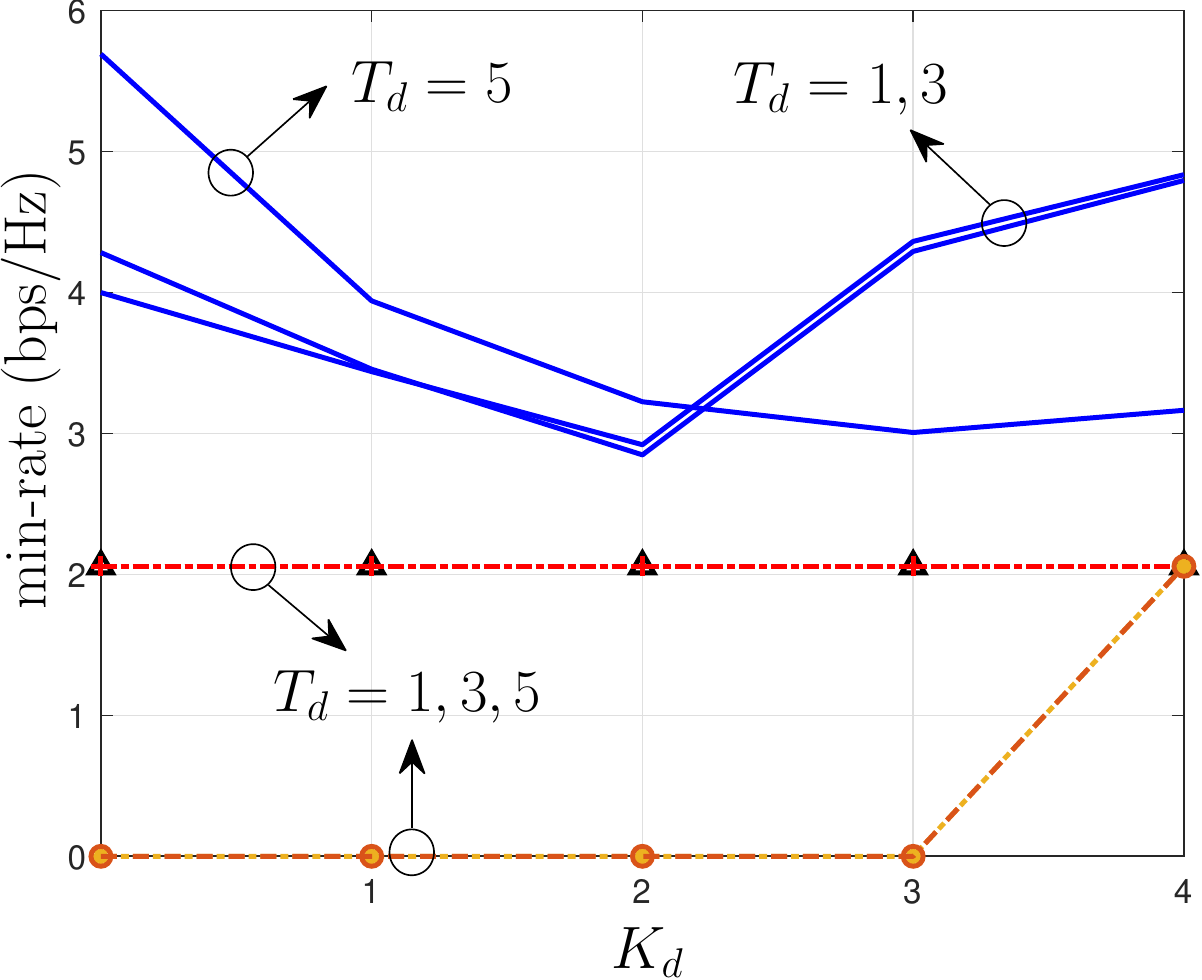}
		\caption{}
		\label{fig:a2_minrate_Nd}
	\end{subfigure}
	~
	\begin{subfigure}[c]{0.64\columnwidth}\centering
		\includegraphics[width=\columnwidth]{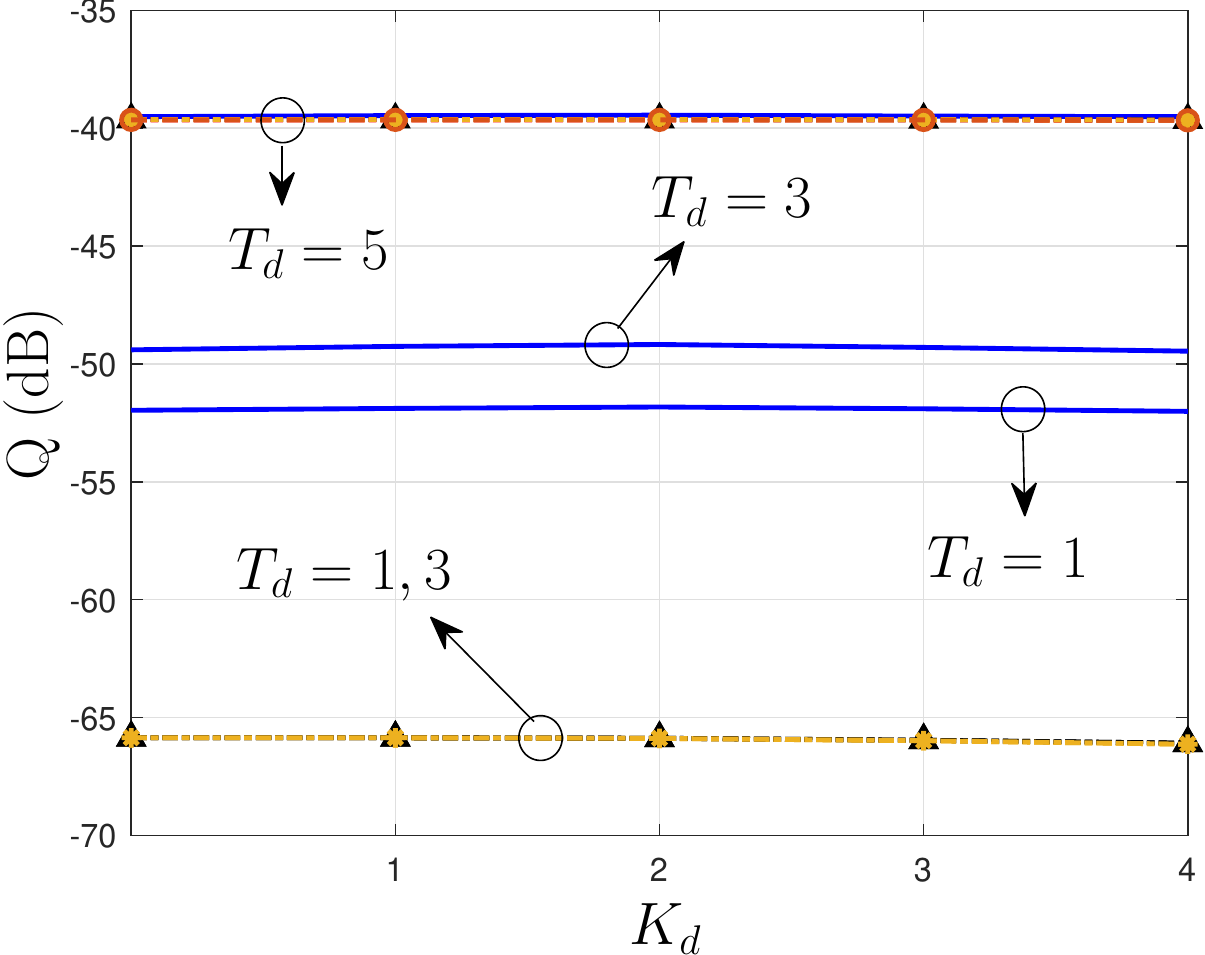}
		\caption{}
		\label{fig:a2_q_Nd}
	\end{subfigure}
	\caption{ Dual \ac{ris}-assisted \acs{isac} system. Impact of $\Gamma$ on (a) fairness \ac{sinr} and (b) worst-case target illumination power. Impact of $N$ on (c) fairness \ac{sinr} and (d) worst-case target illumination power. Impact of $K_d$ on (e) min-rate and (f) worst-case target illumination power.  }
	\label{fig:alg2_sim}
\end{figure*}

	\subsection{Dual RIS-assisted ISAC system} \label{sec:simulations:dual_ris}
	In this subsection, we demonstrate the advantages of using  dedicated \acp{ris} for sensing and communications in an \acs{isac} system. We assume that two identical \ac{ris} with $N$ elements assist the communication and sensing functionalities of the \acs{isac} system. We use the same simulation parameters as before and the parameters are provided in Table~\ref{table:simparam_locations}. Unless otherwise mentioned, we use $K=4$, $T=5$, $\Gamma = 5~{\rm dB}$, $P_t = 0~{\rm dB}$, and $N = 100$. 
	
	Let us recall that the dual RIS-assisted \ac{isac} system is particularly suited for scenarios where some or all of the targets are not directly visible to the \ac{dfbs}. Let $T_d$ denote the number of targets that are directly visible to the \ac{dfbs}. In this section, we present the fairness \ac{sinr} (or min-rate) and the worst-case target illumination power of different schemes for varying simulation parameters such as \ac{sinr} constraints ($\Gamma$), number of \ac{ris} elements~($N$), number of users with direct path $K_d$, and the number of targets that are directly visible to the \ac{dfbs}~($T_d$). All the results in this subsection are obtained by averaging over 100  independent channel realizations with varying user locations.
	
	As before, we compare the performance of the proposed method, i.e., \texttt{RIS proposed}, with that of manually designing beams at each \ac{ris}. Specifically, we select the radar-\ac{ris} (respectively, comm-\ac{ris}) phase shifts to form equally powerful beams towards all the targets locations of interest (respectively, all \acp{ue}). The transmit beamformers are then obtained by solving~\eqref{sp3}. We consider four different scenarios of ISAC systems for comparison: dual RIS with manually selected phase shifts to form interpretable beams, indicated as \texttt{manual selection (both RISs)}; only single radar-RIS  with manually designed phase shifts, indicated as \texttt{manual selection (only radar RIS)}; only single comm-RIS with manually designed phase shifts, indicated as \texttt{manual selection (only comm RIS)}; and \acs{isac} system without RIS, indicated as \texttt{no RIS}. The methods with manual selection of \ac{ris} phase shifts are agnostic to the considered task. In addition, we also use \texttt{sensing-only system}~\cite{stoica2007on_probing_signal} as a baseline system for radar performance.

	In Fig.~\ref{fig:alg2_sim}(a), we present the fairness \ac{sinr} of different methods. The proposed method  clearly outperforms baseline systems based on manual selection of \ac{ris} phase shifts, \acs{isac} system without \ac{ris}~\cite{liu2020joint_transmit_beamform} and sensing-only system~\cite{stoica2007on_probing_signal} while improving the fairness \ac{sinr} by  often more than $15$ dB, irrespective of whether all targets are directly visible to the \ac{dfbs} or not. Since the precoder design phase attempts to maximize $Q$ while satisfying the fairness SINR, the fairness SINR will be about $\Gamma$. This is the reason why all the baseline methods achieve a fairness SINR of $\Gamma$ dB.  However, in the proposed method, once the precoders $\mC$ and $\mS$ are designed, the comm-RIS phase shifts $\boldsymbol{\omega}_{\rm c}$ are also updated so as to maximize the SINR. This leads to a significantly high fairness SINR for the proposed scheme.

	The worst-case target illumination power of different methods is  presented in Fig.~\ref{fig:alg2_sim}(b). As expected, schemes that do not utilize a dedicated radar-\ac{ris} are not able to illuminate all targets efficiently whenever a few targets (say, $T_d = 1,3$) are not directly visible to the \ac{dfbs}. While \texttt{manual selection (only radar RIS)} and \texttt{manual selection (both RISs)} results in a non-zero worst-case target illumination power, the proposed scheme offers more than $12$ dB higher $Q$ than the next best benchmark scheme. As $T_d$ increase from $1$ to $3$, the $Q$ of the proposed method also increase by about 3 dB. This is expected since more number of targets can be now served directly via the \ac{dfbs}. When all targets are visible to the \ac{dfbs} (i.e., $T_d = 5$), the performance of all methods are comparable with the proposed scheme offering a slight improvement. Furthermore, with an increase in the \ac{sinr} requirement, $Q$ of all schemes decrease. This is because of the higher amount of power that needs to be transmitted towards the \acp{ue} to meet the increased fairness \ac{sinr} requirement. 
	
	In Fig.~\ref{fig:alg2_sim}(c) and Fig.~\ref{fig:alg2_sim}(d), we illustrate the impact of number of \ac{ris}  elements ($N$) on the fairness \ac{sinr} and  worst-case target illumination power, respectively. As before, irrespective of the number of targets with direct paths, the proposed scheme offers significant improvements in the fairness \ac{sinr} over other benchmark schemes. Moreover, $Q$ of the proposed method is also remarkably high when compared with schemes where the radar-\ac{ris} phase shifts are manually designed, especially when few of the targets are not directly visible to the \ac{dfbs}. As the number of radar-\ac{ris} (respectively, comm-\ac{ris}) elements increase, the array gain offered by the radar-\ac{ris} (respectively, comm-\ac{ris}) also increase leading to an increase in the worst-case target illumination power (respectively, fairness \ac{sinr}). Since the comm-\ac{ris} is not involved in radar sensing, $Q$ of \texttt{manual selection (only radar RIS)} and \texttt{manual selection (both RISs)} overlap with each other. Similarly, both \texttt{manual selection (only comm RIS)} and \texttt{no RIS} also have identical values of $Q$ because both these scenarios correspond to a setting without the radar-RIS.

	Finally, we show the performance of the proposed dual-RIS-assisted ISAC system in a scenario with the direct paths to a few of the users is blocked in Fig.~\ref{fig:alg2_sim}(e) and Fig.~\ref{fig:alg2_sim}(f). As before, with the use of the comm-\ac{ris}, all users are served with a minimum \ac{sinr} of $5~{\rm dB}$ (i.e., min-rate of $2.05$ bps/Hz). Moreover, with the proposed scheme, all users are served with a higher rate irrespective of $T_d$. As can be observed from Fig.~\ref{fig:alg2_sim}(f), the proposed method results in remarkably improved worst-case target illumination powers, especially when the direct paths to a few of the targets are blocked. In sum, albeit the inability to form uncorrelated beams due to the fully passive nature of \acp{ris}, using an additional \ac{ris} for radar sensing along with the comm-\ac{ris} leads to significant improvements in both radar and communication performance of \acs{isac} systems especially when not all targets are directly visible to the \ac{dfbs}.

	\section{Conclusions} \label{sec:conclusion}
	
	We considered two settings of \ac{ris}-assisted \ac{isac} systems, namely, a setting with a single \ac{ris} that assists only the communication functionality of an \ac{isac} system and a setting with two \acp{ris} wherein dedicated \acp{ris} are used for sensing and communications. We developed algorithms to design the transmit beamformers to jointly precode communication symbols and radar waveforms and to also design RIS phase shifts to achieve certain sensing performance in terms of the beampattern matching error or worst-case target illumination power while ensuring a minimum SINR for the communication \acp{ue}. Since the resulting optimization problems are non-convex, we have developed alternating optimization solvers for the design problems that appear in the two settings. A semi-definite convex problem is used to solve for transmit beamformers with fixed RIS phase shifts. With fixed transmit beamformers, the design of comm-RIS phase shifts is carried out using generalized Dinkelbach iterations. With fixed transmit beamformers, the design of radar-RIS phase shifts is carried out using SDP followed by Gaussian randomization. The performance of the proposed algorithms is then demonstrated through numerical simulations. Specifically, the comm-RIS assisted ISAC system is found to significantly improve the fairness SINR while suffering from a moderate loss in radar performance. On the other hand, the dual-RIS assisted ISAC system is found to remarkably improve both communication and radar performance metrics, especially in scenarios where all the targets are not directly visible to the \ac{dfbs}.

	\bibliographystyle{IEEEtran}
	\bibliography{IEEEabrv,bibliography}

\end{document}